\documentclass[10pt, preprint]{aastex}
\usepackage{epsfig}
\usepackage{changebar}
\usepackage{emulateapj5}
\usepackage{graphicx}

\bibpunct{(}{)}{;}{a}{}{,}

\slugcomment{To appear in ApJ}

\shorttitle{ACBAR Power Spectrum}
\shortauthors{Kuo et al.}

\usepackage{amsmath}
\usepackage{amssymb}

\begin{document}

\title{Improved Measurements of the CMB Power Spectrum with ACBAR}


\author{C.L. Kuo\altaffilmark{1,2}, 
P.A.R. Ade\altaffilmark{3},
J.J. Bock\altaffilmark{1,2}, 
J.R. Bond\altaffilmark{4},
C.R. Contaldi\altaffilmark{4,5},
M.D. Daub \altaffilmark{6}, 
J.H. Goldstein\altaffilmark{7}, 
W.L. Holzapfel \altaffilmark{6}, 
A.E. Lange\altaffilmark{1,2}, 
M. Lueker\altaffilmark{6},
M. Newcomb\altaffilmark{8},
J.B. Peterson\altaffilmark{9},
C. Reichardt\altaffilmark{1},
J. Ruhl\altaffilmark{7}, 
M.C. Runyan\altaffilmark{10},
Z. Staniszweski\altaffilmark{7}}

\altaffiltext{1}{Observational Cosmology, California Institute of Technology, MS 59-33, Pasadena, CA 91125}
\altaffiltext{2}{Jet Propulsion Laboratory, 4800 Oak Grove Drive, Pasadena, CA 91109}
\altaffiltext{3}{Department of Physics and Astronomy, Cardiff University, CF24 3YB Wales, UK}
\altaffiltext{4}{Canadian Institute of Theoretical Astrophysics, University of Toronto, Toronto, Ontario, M5S3H8,
Canada}
\altaffiltext{5}{Blackett Laboratory, Imperial College, Prince Consort Road, London, SW7 2AZ, U.K.}
\altaffiltext{6}{Department of Physics, University of California, Berkeley, CA 94720}
\altaffiltext{7}{Department of Physics, Case Western Reserve University, Cleveland, OH 44106}
\altaffiltext{8}{Yerkes Observatory, 373 W. Geneva Street, Williams Bay, WI 53191}
\altaffiltext{9}{Department of Physics, Carnegie Mellon University, Pittsburgh, PA 15213}
\altaffiltext{10}{Department of Physics, University of Chicago, Chicago IL 60637}

\begin{abstract}

We report improved measurements of temperature anisotropies in the
cosmic microwave background (CMB) radiation made with the Arcminute Cosmology
Bolometer Array Receiver (ACBAR).  In this paper, we use a new
analysis technique and include 30\% more data from the 2001 and 2002
observing seasons than the first release~\citep{kuo04} to derive a new
set of band-power measurements with significantly smaller
uncertainties.  The planet-based calibration used previously has been
replaced by comparing the flux of RCW38 as measured by ACBAR and
BOOMERANG to transfer the WMAP-based BOOMERANG calibration to ACBAR.
The resulting power spectrum is consistent with the theoretical
predictions for a spatially flat, dark energy dominated $\Lambda$CDM
cosmology including the effects of gravitational lensing.  
Despite the exponential damping on small angular scales,
the primary CMB fluctuations are detected with a signal-to-noise ratio
of greater than $4$ up to multipoles of $\ell=2000$.  
This increase in
the precision of the fine-scale CMB power spectrum leads to only a
modest decrease in the uncertainties on the parameters of the standard
cosmological model.
At high angular resolution, secondary anisotropies are predicted to be a
significant contribution to the measured anisotropy.
A joint analysis of the ACBAR results at
$150\,$GHz and the CBI results at $30\,$GHz in the multipole range
$2000 < \ell < 3000$ shows that the power, reported by CBI in excess of the predicted 
primary anisotropy, has a frequency spectrum consistent with
the thermal Sunyaev-Zel'dovich effect and inconsistent with primary CMB. 
The results reported here are derived from a subset of the 
total ACBAR data set;
the final {\sc ACBAR} power spectrum at $150\,$GHz will include 3.7 times more 
effective integration time and 6.5 times more sky coverage than is used here.

\end{abstract}


\section{Introduction}\label{sec:intro}

Observations of the cosmic microwave background (CMB) radiation on 
angular scales corresponding to multipole values of $\ell \lesssim 1000$ 
have established a ``concordance" 
cosmological model characterized by a negligible spatial curvature, 5\% 
baryonic matter, 25\% dark matter, and  70\% dark 
energy \citep{spergel06}, in good agreement 
with results from other cosmic probes 
\citep{tegmark04,riess04,latestbbn}. 
On smaller angular scales, the CMB anisotropy is exponentially damped by 
photon diffusion. The damping scale is a measure 
of the angular size of the Silk length at the surface of last scattering 
\citep{silk68,bond84,hu97}. This portion of the CMB power spectrum, known as 
the ``damping tail'', provides a consistency check of the cosmological
model and independent constraints on several cosmological parameters.  

At $150\,$GHz, anisotropies from the thermal 
Sunyaev-Zel'dovich (SZ) effect are expected to dominate over the 
primary CMB fluctuations for multipoles $\ell > 2500$.  
The fluctuation power of this ``SZ excess" depends sensitively on the 
integrated cluster abundance and cluster gas distribution. 
Accurate measurements of the SZ fluctuation amplitude provide 
an independent measurement of the amplitude of matter perturbations,
usually characterized by $\sigma_8$. 

The Arcminute Cosmology Bolometer Array Receiver (ACBAR) is 
designed to study both primary and secondary CMB anisotropies
on small angular scales \citep{runyan03a}.
The fundamental features of the angular power spectrum in the range 
$\ell \lesssim 1000$ have 
been characterized by several ground, balloon, and satellite-based experiments.
The damping of CMB anisotropy power at $\ell \gtrsim 1000$ has been measured 
by the Cosmic Background Imager (CBI) 
\citep{pearson03,readhead04}, ACBAR, and the Very 
Small Array (VSA) \citep{dickinson04}. 
The excellent agreement between the observed CMB damping tail power spectrum 
and the theoretical predictions of the $\Lambda$CDM model provide compelling
evidence that our interpretation of the CMB is correct.  

The high resolution CMB anisotropy measurements reported here
extend the low-$\ell$ results, such as those from the WMAP satellite, 
to a comoving scale of $\sim 5$~Mpc, and provide strong constraints 
on the shape of the primordial density perturbation
spectrum~\citep{peiris03,mukherjee03,bridle03,leach03}.  
The primordial spectrum is close to scale-invariant, consistent with
predictions of slow-roll inflation models.  
Departures from scale-invariance provide valuable clues to the 
physics of the early Universe. 

Measurements of the high-$\ell$ CMB anisotropy
power at $30\,$GHz by CBI~\citep{mason03,bond05} and BIMA~\citep{dawson06} 
detect power in excess of the predictions of the standard cosmological model.
If interpreted as the thermal SZ effect produced by clusters of galaxies, 
this excess power corresponds to a value of $\sigma_8$ that is slightly 
higher than the value found
by analysis of optical galaxy clustering \citep{tegmark04,2df}.
This result hinges upon the excess fluctuation power seen by the CBI,
which could be the result of contamination by a population of low-flux 
flat-spectrum radio sources. This possibility is being investigated 
with a $30\,$GHz continuum receiver at the Green Bank Telescope \citep{mason}. 
Alternative origins of the excess involve non-standard ingredients in the 
cosmological model, such as, primordial voids and magnetic 
fields \citep{voids,Bfields}.
The frequency spectrum of the thermal SZ effect is distinctly 
different from that of the primordial anisotropy.
ACBAR observes at a much higher frequency than BIMA or the CBI 
($150\,$GHz versus $30\,$GHz), and the addition of the ACBAR high-$\ell$ 
measurements can be used to constrain the possible origin of any 
observed excess.

The first ACBAR power spectrum, presented by \cite{kuo04} (K04), was 
produced from the Lead-Main-Trail (LMT) analysis of a subset of the 
first two years of ACBAR observations.    
Three major improvements are made in this work over that first 
release. 
First and most importantly, the power spectrum is 
derived from the un-differenced temperature maps (rather than 
``Lead-Main-Trail" subtracted maps), significantly 
reducing the uncertainties from cosmic variance and instrumental noise.
Second, we replace the planet-based calibration with a more precise 
calibration based on the flux of RCW38 as measured by ACBAR and 
BOOMERANG \citep{crill03}.
Third, we include two additional CMB fields observed late in the 2002 season
which were not used in previous publications.  
These fields increase the total data volume by $\sim 30\%$.  

This paper is organized as follows. In \S~\ref{sec:instrument} we
review the ACBAR instrument and the CMB observation program.
The new analysis algorithm for the un-differenced maps is explained 
in \S~\ref{sec:analysis}. Section \S~\ref{sec:calib} is an 
overview of the calibration using RCW38;
the details of cross-calibration between 
BOOMERANG and ACBAR are discussed in Appendix A. 
Systematic tests and foreground contamination are discussed in 
\S~\ref{sec:sys}. 
We present the band-power 
results in \S~\ref{sec:results}, including a discussion of the scientific
interpretation. 
The ACBAR band powers are combined with the results of other experiments to place 
constraints on the parameters of cosmological models in 
\S~\ref{sec:parameters}.
The main results of this paper are summarized in 
\S~\ref{sec:conclusion}.

\section{The Instrument And Observations}\label{sec:instrument}
The ACBAR instrument was designed to be used with the Viper telescope 
at the South Pole to observe CMB temperature anisotropies with an 
angular resolution of $5^{\prime}$.  The receiver consists of 16 bolometers, coupled to the 
2-meter off-axis Gregorian telescope through corrugated feed horns.
The bolometers are cooled to 240 mK by a 3-stage He$^3$-He$^3$-He$^4$
sorption refrigerator.  
The beams from the array are swept across the sky at near-constant elevation
by the motion of a flat tertiary mirror.  
The receiver was deployed in December 2000, and
CMB data were taken during the austral winters of 2001, 2002, 2004, and 2005.
The bolometer array was reconfigured between the 2001 and 2002 observing 
seasons to double the number of $150\,$GHz detectors; 
details of the instrument configuration and 
performance in each season are given in \citet{runyan03a}, while details of the 
CMB observations, data reduction procedures, and beam maps can be found in K04.

The results reported here are derived from the
$150\,$GHz data gathered in the  2001 and 2002 Austral winters.
These data come from observations of 
four independent CMB fields, detailed in Table~\ref{tab:fields}.  
The power spectrum derived from the first two fields, CMB2/CMB4 and CMB5,
was reported in K04.  Since then, we have completed the analysis of  
two additional fields, CMB6 and CMB7, observed in 
July and August of 2002. 
Each of these four widely-separated fields was chosen to include a bright
quasar, the image of which provides an important check of the 
telescope pointing over the entire observation period. 

\begin{table*}
\begin{center}
\caption{CMB Fields}
\small
\begin{tabular}{ccccccc}
\hline\hline
\rule[-2mm]{0mm}{6mm}
Field & RA (deg) & dec (deg) & Area (deg$^2$) & Time (hrs.) &Year&\# of detectors\\
\hline
CMB2(CMB4) & 73.963 & -46.268 & 44(28) & 506(142) &2001(2002)&4(8)\\
CMB5 & 43.372 & -54.698 & 31 & 1656 &2002&8\\
CMB6 & 32.693 & -50.983 & 29 & 351 &2002&8\\
CMB7 & 338.805 & -48.600 & 32 & 420 &2002&8\\
\hline
\end{tabular}
\tablecomments{\small
The central quasar coordinates and size of each CMB field observed by ACBAR
during 2001 and 2002.  The fifth column gives the detector integration time
for each field after cuts.  This represents approximately 50\% of the total
time spent observing CMB fields. The last column gives the number of $150\,$GHz detectors.}
\label{tab:fields}
\normalsize
\end{center}
\end{table*}

We derive the noise properties of the data from the raw time streams.
The photon noise and bolometer noise are white, Gaussian distributed, and
uncorrelated between bolometers.  On the other hand, 
``sky noise'' associated with atmospheric fluctuations can introduce
correlations between bolometers. This noise component can be described by 
the Kolmogorov-Taylor theory, which models 
the turbulence as a screen of frozen fluctuations blown through 
the field of view with a certain wind speed.  The sky noise 
properties of the South Pole characterized with ACBAR are reported 
in detail by \citet{bussmann05}.
In the CMB power spectrum analysis, we calculate the 
full cross-channel correlations and use them as a data cut.
We disregard data that show cross-channel 
correlations (after removal of up to a 10th order polynomial from 
each chopper sweep) higher than 5\% of the total noise power.

\section{Un-differenced Power Spectrum Analysis}\label{sec:analysis}

Following the conventions of the first data release,
the band-powers ${\bf q}$ are reported in units of $\mu K^2$, 
and are used to parameterize the power spectrum according to
\begin{equation}
\ell(\ell+1)C_{\ell}/2\pi\equiv{\cal D}_{\ell}=\sum_Bq_B\chi_{B\ell}\;,\label{dl}
\end{equation}
where $\chi_{B\ell}$ are tophat functions; $\chi_{B\ell}=1$ for $\ell \in B$, and
$\chi_{B\ell}=0$ for $\ell \not\in B$.
The observations in 2001 and 2002 were carried out in a {\em lead-main-trail}
(LMT) pattern in order to facilitate the removal of
slowly varying time-dependent 
chopper synchronous offsets.
In the analysis of K04, we differenced the CMB maps 
according to the formula $M-(L+T)/2$, and derived the band-powers from the
differenced maps.  While this conservative strategy can prevent potential
systematic errors, it increases the effect of instrumental noise and reduces 
sky information. 
At high-$\ell$, where the uncertainties in the band-powers are 
dominated by the instrument noise,
the anisotropy power signal-to-noise ratio (S/N) 
derived from a LMT differenced map is a factor of
$$
\eta_{LMT}=\frac{SN_{LMT}}{SN_0}=
$$
$$
\left[
\frac{\sqrt{1^2+(1/2)^2+(1/2)^2}}{\sqrt{\sqrt{2}^2+(\sqrt{4}/2)^2+(\sqrt{4}/2)^2}}\right]^2
=\frac{3}{8}
$$
smaller than that for an un-differenced map with an identical amount of observing 
time (ignoring the small correlations between differenced fields). The higher noise in
L and T reflects the reduced (50\%) observing time on these fields compared to the main field.

It is possible to generalize the noise-weighted co-added map analysis 
outlined in K04 to avoid this S/N degradation, while ensuring that the 
power spectrum is not contaminated by the chopper 
synchronous signals. The key to this generalization lies in the fact that the
band-powers can be derived from arbitrary linear combinations
of the time stream.  The linear combinations can be constructed such that
the un-differenced maps are cleaned of chopper synchronous
offsets. The results presented in \S~\ref{sec:results} are derived from 
such ``cleaned'', un-differenced maps, using the method outlined below. 

Suppose $d_\alpha\;$($\alpha=1..n_t$) are $n_t$ time-ordered measurements of CMB
temperature. This vector is the sum of the noise component
$n_\alpha$ and the signal component $s_\alpha$. The data from each chopper sweep
are filtered to remove chopper synchronous offsets
before they are co-added.
For example, since the beams approximately follow a great circle
on the celestial sphere, changing the chopper angle causes the detectors to 
look through a different atmosphere thickness and produces
signals that are functions of the chopper position.
Formally, the filtering is done by operating on the time stream {\bf d} 
with the ``corrupted mode projection'' matrix
${\bf \Pi}$. 
This operation results in the cleaned
time stream ${\tilde{\bf d}}\equiv {\bf \Pi d}$. 

Large angular scale chopper offsets are largely removed by this mode projection. 
In the analysis of K04, any residual small angular scale offsets were eliminated 
through LMT differencing. 
Alternatively, chopper synchronous offsets can be removed by
subtracting the average chopper function, determined with the whole map, 
from the individual data strips \citep{runyan_the}.
However, this assumes that the offsets do not change in time and elevation. 
In this paper, we remove a chopper synchronous offset from each data strip where 
the amplitude of the offset at each sample in the strip is free to vary 
quadratically with elevation in the map.
The mean of these quadratic functions in elevation gives the mean chopper function; 
at zeroth order, the mean chopper function is removed.
The quadratic variation allows for slow changes in the synchronous offset as a 
function of time and elevation.  
Projecting out these corrupted modes eliminates any detectable residual 
chopper offset with minimal loss of signal in the final power spectrum. 
The loss of information at high-$\ell$ is negligible, since the
data contain many more degrees of freedom than the removed modes do.

Mathematically, the corrupted mode projection matrix $\Pi$
is now the product of two matrices, $\Pi\equiv\Pi_2 \Pi_1$.
The operator $\Pi_1$ is the original $\Pi$ matrix used by K04 that 
adaptively removes polynomial modes in RA, 
where the order of the polynomial removed typically depends on the atmospheric 
conditions. 
The additional operator $\Pi_2$ removes quadratic modes in DEC independently 
for each of the lead, main, and trail fields. 
We perform systematic tests in \S\ref{sec:sys} to confirm that any residual 
offset is smaller than the noise level of the final power spectrum.
The resulting time stream ${\tilde{\bf d}}$ is then co-added
into a map ${\bf T}$ according to the telescope pointing model.
Next, a signal-to-noise eigenmode truncation is applied, 
and only the modes expected to have non-negligible 
signal-to-noise ratio are retained. This significantly reduces the computational 
requirements of the analysis. Since both the projection of corrupted 
modes and signal-to-noise eigenmode truncation are linear operations, 
the resulting data vector ${\bf \Delta}$ is also a linear combination
of the original time stream, represented by
$$
{\bf \Delta}={\bf L}{\bf d}.
$$
The noise covariance matrix is given by
$$
{\bf C}_N={\bf L}\langle {\bf n n}^t \rangle {\bf L}^t.
$$
The signal component in the time stream ${\bf s}$ is the convolution 
of sky map ${\mathfrak T}({\bf r})$ with $B_\alpha({\bf r})$, the beam 
function  during measurement 
$\alpha$,
$$
s_\alpha=\int d^2r {\mathfrak T}({\bf r}) B_\alpha({\bf r}).
$$
The signal component of $\Delta$ is 
\begin{equation}
\Delta_i^{sig}=\sum_\alpha L_{i\alpha}s_\alpha\equiv\int d^2r F_i({\bf r})
{\mathfrak T}({\bf r}),\label{tsig}
\end{equation}
where the pixel-beam function $F_i$ is given by
\begin{equation}
F_i({\bf r})=\sum_\alpha L_{i\alpha}B_\alpha({\bf r}).
\label{qbeam}
\end{equation}
In the flat sky limit, the theory covariance matrix is given by
$$
C_{T\{ij\}}=
\langle \Delta_i\Delta_j\rangle^{sig}=
\int\!\!\int d^2r d^2r' F_i({\bf r})F_j({\bf r}')
\langle {\mathfrak T}({\bf r}){\mathfrak T}({\bf r}')\rangle
$$
$$
=\int\!\!\int d^2r d^2r' F_i({\bf r})F_j({\bf r}') \int \frac{d^2l}{(2\pi)^2}
C_{\ell} \cdot e^{i{\bf l}\cdot ({\bf r}-{\bf r}')}
$$
\begin{equation}
=\int \frac{d^2\ell}{(2\pi)^2}C_{\ell} \cdot {\tilde F}_i^*({\bf l}){\tilde F}_j({\bf l}),
\label{ct} 
\end{equation}
where ${\tilde F}_i({\bf l})$ is the Fourier transform of $F_i({\bf r})$.
To perform the iterative quadratic band-power estimation procedure,
it is necessary to know the partial derivative of $C_T$ with respect to 
each of the band-powers $q_B$, which according to equation(\ref{dl}) is given by
\begin{equation}
\frac{\partial C_{T\{ij\}}}{\partial q_B}=\int \frac{d^2\ell}{(2\pi)^2} 
\frac{2\pi\chi_{B\ell}}{\ell(\ell+1)}
\cdot {\tilde F}_i^*({\bf l}){\tilde F}_j({\bf l}).\label{ctp}
\end{equation}
Note that this algorithm does not assume that the instrument beams 
stay constant during the observations. As described in 
\citet{runyan03a}, the ACBAR beam sizes are weak functions of the chopper 
position. K04 adopted a semi-analytic expansion to correct
for these effects to first order. To verify that the effects due to non-uniform 
beams are small, we developed two end-to-end pipelines. In the first pipeline,
the pixel-beam functions $F_i({\bf r})$ are calculated explicitly during the co-adding process. The bandpowers in Table \ref{tab:bands} are analyzed with this algorithm. In the second pipeline, an averaged beam is used for the entire map. The difference in the power spectra from the two pipelines is negligible.

In the analysis of K04, we assumed that the noise is stationary 
in chopper position after LMT subtraction. In the current treatment, we 
relax this assumption and calculate the full two dimensional
correlation matrix directly from the time stream data without using 
Fourier transforms.

\begin{figure*}[ht!]
\resizebox{\hsize}{!}{
\plotone{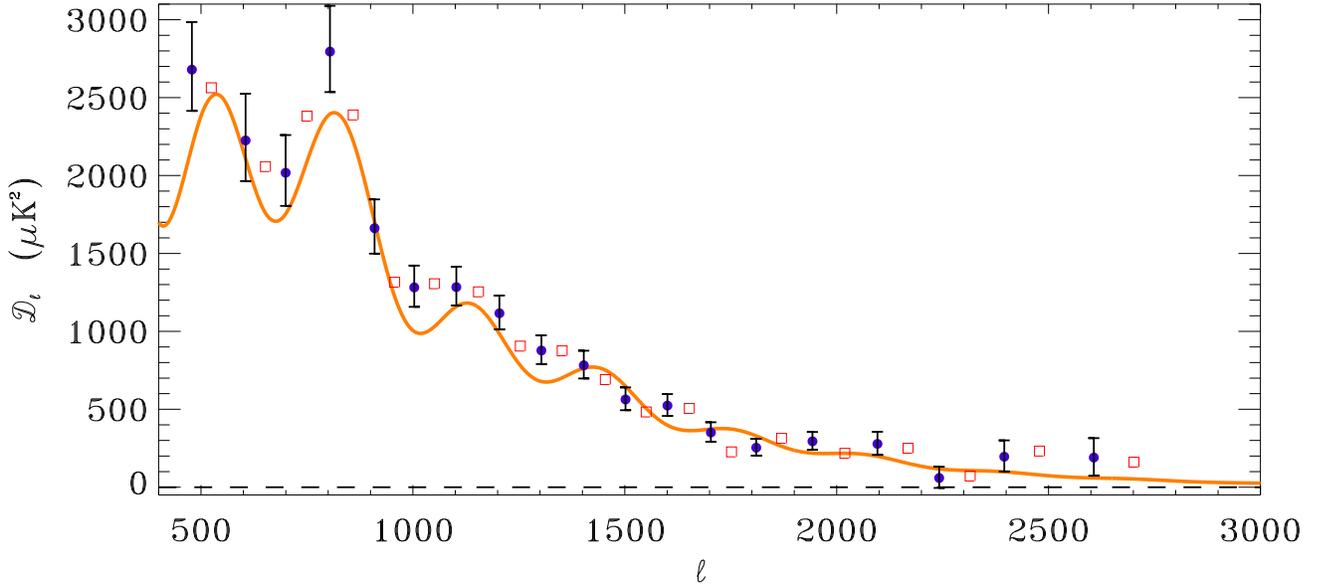}}
\caption{
The de-correlated ACBAR band-powers for two alternate binnings. These two binnings are
not independent, therefore only one set is shown with error bars, which
correspond to 1-$\sigma$ uncertainties calculated from the offset lognormal
fits to the likelihood function. 
Both the overall features and the damping scale are in good agreement with predictions
from a flat, low baryon density $\Lambda$CDM Universe.
The third acoustic peak (around $\ell=800$) is clearly seen. Small scale primary 
CMB fluctuations are detected with high signal-to-noise ratio 
($>4$) up to $l=2000$. The plotted model line is the best fit to the WMAP3 and ACBAR
bandpowers.}
\label{fig:acbar} 
\end{figure*}

All the numerical calculations are performed on the 
National Energy Research Scientific Computing Center (NERSC) IBM SP RS/6000. 
The evaluation of $F_i({\bf r})$ and its Fourier transform
are the most computationally demanding steps in this analysis. 
After ${\bf C}_T$, ${\bf C}_N$ and ${\bf C}_{T,B}$ are calculated,
standard likelihood maximizing procedures are used to find the 
band-powers ${\bf q}_B$ and uncertainties \citep{bond98}. 
The results of this analysis are presented in Table \ref{tab:bands} and 
Figure \ref{fig:acbar}.

\section{Calibration}\label{sec:calib}

RCW38 is a compact HII region in the Galactic plane at a declination similar to the 
ACBAR CMB fields. 
It has a large and stable flux and serves as the primary calibrator for the 
ACBAR observations.  
We determine the absolute flux of RCW38 using maps from the 2003 flight of BOOMERANG 
(\cite{masi05}, hereafter B03),
which are calibrated relative to the WMAP experiment with an absolute uncertainty of $1.8\%$.  
RCW38 does not have a black-body spectrum, requiring spectral corrections for the
calibration of CMB anisotropies. 
However the similarity in the spectral responses of the $150\,$GHz 
bands in the B03 and ACBAR experiments ensures these corrections to be small. 
Here we outline the calibration procedure, leaving the details to Appendix \ref{app:calib}.

ACBAR typically observed RCW38 before and after each CMB observation. 
Comparisons between the B03 and ACBAR maps of RCW38 are used to 
determine the absolute calibration of the CMB fields to an uncertainty of 6.0\%. 
For roughly 50\% of the 2002 season, we observed RCW38 with only half the $150\,$GHz 
detectors (4 out of 8).  During these periods, the RCW38 calibration was applied to 
the remaining detectors by comparing CMB power spectra derived from each half of 
the detectors.  
The calibration of the CMB4 field (observed in 2002) is extended to the 70\% overlapping 
CMB2 field (observed in 2001) by comparing power spectra from each field.  
In the first ACBAR release, the 2001 and 2002 data sets were calibrated with an 
accuracy of 10\% using observations of Mars and Venus respectively. 
We determine the corrections to this planet-based calibration to be 
0.911$\pm$0.072 for CMB2 (2001) and 1.128$\pm$0.066 for CMB4-7 (2002) in CMB temperature. 
The 2002 observations dominate the final power spectra, and 
the final results have essentially the same 6\% temperature calibration 
uncertainty as the 2002 data.

\section{Systematic Uncertainties and Foregrounds}\label{sec:sys}

\subsection{Jackknife Tests}
\begin{figure*}[ht]
\centering
\includegraphics[width=4.5in]{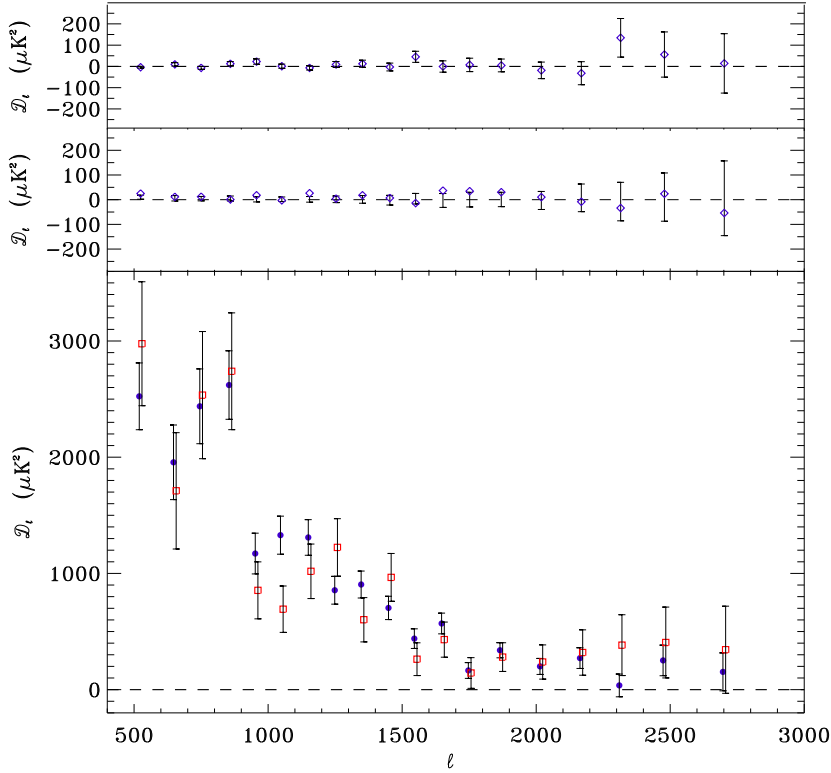}
\caption{Systematic tests performed on the ACBAR data. {\em Top}:
Power spectrum produced from the difference of maps made with left-
and right-going chopper sweeps.
 {\em Middle}: Power spectrum (diamonds) calculated
from the difference of maps made from the first and second halves of the data for each field
(CMB2, CMB5, CMB6, and CMB7), compared with Monte Carlo simulations (error bars).
{\em Bottom}: Power spectrum calculated from the LMT sum maps (squares), compared with the
joint power spectrum (pre-decorrelated, filled circles).
The consistency between two sets of
band-powers demonstrates that the residual chopper offsets are below noise.
See text for more detail on this test.}\label{fig:sys}
\end{figure*}

We performed a series of tests to constrain 
the amplitude of potential systematic errors in the power spectrum results.
As described by K04, the ``first half minus second half" 
jackknife is a very powerful test for time dependent
errors, such as a changing 
calibration, inconsistency 
in the beam or pointing reconstruction, and time varying sidelobe pickup.
In addition, high-$\ell$ jackknife band-powers constrain the mis-estimation 
of noise.
We perform this test on the joint CMB power spectrum and
find the band-powers of the chronologically differenced maps 
are consistent with zero (Fig. \ref{fig:sys}). 

Similarly, the data can be divided in two halves according to the 
direction of the chopper motion. Microphonic vibrations due to the chopper
turn-arounds, erroneous transfer function corrections, or effects of wind 
direction could produce a nonvanishing signal in the jackknife band-powers.
We find that the power spectrum of the left-right differenced maps
is also consistent with zero.

To ensure that any residual chopper synchronous offset is below the noise 
level, we have developed a new systematic test that {\em enhances} the 
contribution of any such offsets relative to the CMB.
In this test, the band-powers are derived
from an LMT {\em sum} map, $L+M+T$, in which the residual 
synchronous offsets (the same in each field) are enhanced relative 
to the (random) CMB fluctuations by a factor of 
3 in power (neglecting the small correlations at low-$\ell$). The resulting 
band-powers are compared with the un-differenced band-powers to 
check for systematic deviations.
This test is particularly sensitive to any residual chopper offsets.
We find no significant deviation 
in the LMT sum band-powers from the un-differenced band-powers. 
When compared with the model $\Lambda$CDM power spectrum, we do notice a slight rise in the 
LMT sum band-powers for $\ell>2300$. It is difficult to assess the significance 
or the origin of this low level trend. However, even if it is caused by 
a residual systematic effect, the contribution to the joint band-powers 
would be smaller than the statistical uncertainty in the reported band powers 
in this paper after accounting for the factor of 3 amplification. 

\subsection{Foregrounds}

At frequencies below the peak of the CMB intensity ($\sim 200\,$GHz), the contribution of 
extra-galactic radio point sources to the observed CMB temperature anisotropy 
decreases rapidly with increasing observing frequency. 
In addition to the negative spectral indices of the majority of the radio sources,
the flux-to-temperature conversion factor, $(dB_\nu/dT_{CMB})^{-1}$,
reaches a minimum as the observing frequency approaches the peak of the CMB. 
In particular, this factor is nearly 15 times smaller at $150\,$GHz than at $30\,$GHz.   
The measurements of ACBAR are therefore much less susceptible to contamination 
by radio point sources than experiments operating at $30\,$GHz such as CBI, BIMA, and VSA.   
We construct templates using the positions of the known radio sources
from the $4.85\,$GHz Parkes-MIT-NRAO (PMN) survey \citep{wright94}, and project out their
contributions to the power spectrum estimations. Using the method described by 
K04, we remove them from the data without making assumptions about their 
fluxes. 
Of 200 PMN sources in the observed CMB fields, we detected the guiding quasars
and six additional sources with significance greater than $>2.8\sigma$. 
These sources tend to have shallow, and in some cases inverted
spectral indices. Table \ref{tab:pmnsources} lists the parameters of the PMN sources that are 
detected in the ACBAR fields; the detection threshold of $>2.8\sigma$, corresponds to a 
false detection rate of 1. The uncertainties are calculated using
Monte-Carlo simulations, and are dominated by contributions from CMB primary 
anisotropies. 
With the exception of the guiding quasar in each of the CMB fields,
the effect of removing the point sources on the band-powers is not significant.

Thermal emission from interstellar dust also has the potential to 
contaminate the measured power spectrum. 
The ACBAR CMB fields are located in the regions of low Galactic 
dust emission. \citet{finkbeiner99} (FSD) combined observations from 
IRAS, COBE/DIRBE, and COBE/FIRAS to generate a multi-component dust model
that predicts the thermal emission at CMB frequencies
with an angular resolution of 6 arcminutes. We apply the ACBAR filtering
to the predicted dust maps for $150\,$GHz, and find the expected RMS to be at the 
$\mu K$ level. Assuming the ACBAR maps contain the FSD dust templates
with amplitudes parametrized by the quantity $\xi$, the observed maps 
$T$ can be written as the sum $T_{CMB}+\xi T_{FSD}$. 
After cross-correlating the dust template maps $T_{FSD}$ with 
the observed maps, we find that the Galactic dust 
is undetectable in the ACBAR $150\,$GHz data. The 1-$\sigma$ upper limit on the amplitude parameter
is $\xi< 2.6$, consistent with the FSD predictions ({\em i.e.}, $\xi=1$). 
As in the case for the radio source flux measurements, the uncertainty 
in $\xi$ is dominated by the CMB primary anisotropies. 
Therefore, dust with the morphology of the FSD maps does not significantly contribute to
the observed anisotropy power.
Nonetheless, the reported band-powers are calculated with the dust template mode 
projected out in each of the fields.

Dust emission from high redshift star forming galaxies can be a significant  
foreground contaminant in millimeter wavelength CMB maps. 
Recent observations with SCUBA/JCMT \citep{smail02,chapman02,borys03}, 
Bolocam/CSO \citep{laurent05}, and MAMBO/IRAM \citep{greve04}, 
provide constraints on both the source counts and the spectral dependence of these 
proto-galaxies. 
However, despite the tremendous progress made in studying these sources, 
their contributions at $150\,$GHz
are still highly uncertain. The uncertainties come from the low number statistics in source counts and 
spectral dependence, difficulties in modeling the survey bias and completeness, 
and poorly studied angular correlations.
In the absence of more decisive measurements, such as might be produced by the ongoing
SHADES\footnote{http://www.roe.ac.uk/ifa/shades/} survey and 
the BLAST\footnote{http://chile1.physics.upenn.edu/blastpublic/} 
experiment, we ignore the clustering 
noise component and estimate the Poisson contribution from these 
proto-galaxies at the ACBAR observing frequencies.

The SCUBA results constrain the $850\mu$m source counts with an uncertainty of 
$\sim 40\%$. There are considerable uncertainties in extrapolating this result to lower 
frequencies because the process depends not only on the dust properties, 
but also on the cosmic star formation history and the source evolution. 
Observations carried out at two different wavelengths, $1.2\,$mm with MAMBO, and 
$1.1\,$mm with Bolocam, can potentially provide this extrapolation phenomenologically.
\citet{greve04} find that the MAMBO and SCUBA source counts agree if the 
MAMBO counts are 
scaled up in flux by a factor of 2.5, corresponding to a spectral dependence of
$S_\nu\propto \nu^{2.65}$. 
We scale the fit for the SCUBA source count results \citep{borys03} to 
$150\,$GHz ($2\,$mm) according this spectral dependence. 
Using the formulae given in \citet{scott99}, we find the contribution of these sources 
to the CMB power spectrum at $150\,$GHz to be ${\cal D}_{\ell}\sim 37(\ell/2500)^2 \mu {\rm K}^2$. 
Since most of this fluctuation power comes from sources with fluxes between $0.1$ and 
$10\,$mJy, the result is insensitive to a flux cut-off greater than $10\,$mJy.
The Bolocam team \citep{laurent05} found fewer sources at $1.1\,$mm than MAMBO did at 
$1.2\,$mm, implying a steeper spectrum for the sources. 
A fluctuation analysis of the same data
also suggested lower source counts, especially for fluxes $< 1$mJy \citep{maloney05}. 
From Figure 15 of \citet{laurent05}, we estimate the spectra between 
850 $\mu$m and $1.1\,$mm to scale as $S_\nu\propto \nu^{4}$. 
Compared with the $37\mu{\rm K}^2$ at $\ell=2500$ from the MAMBO/SCUBA extrapolation, which is just below the 
instrumental noise of ACBAR, the Bolocam/SCUBA model results in a negligible value of $4 \mu{\rm K}^2$.
Future sub-millimeter and millimeter observations are needed to fully characterize the
properties of these sources. 
In the interpretation of the ACBAR data, we cautiously assume that the high-$\ell$ band-powers are not 
significantly contaminated by high redshift proto-galaxies.

\begin{table*}
\begin{center}
\caption{Millimeter Bright PMN Sources}
\small
\begin{tabular}{ccccc}
\hline\hline
\rule[-2mm]{0mm}{6mm}
Source Name/Position & Field & $S_{4.85}$ (mJy)& $S_{150}$ (mJy)&$\alpha_{150/4.85}$\\
\hline
PMN J0455-4616$^*$& CMB2 &$1653$&$2898\pm 60$&0.15\\
PMN J0451-4653 & CMB2 & $541$ & $360\pm 58$&-0.13\\
PMN J0439-4522 & CMB2 & $634$ & $383\pm 73$&-0.16\\
PMN J0514-4554 & CMB2 & $422$ & $197\pm 68$&-0.23\\
PMN J0515-4556 & CMB2 & $990$ & $524\pm 65$&-0.20\\
PMN J0253-5441$^*$& CMB5 & $1193$&$1799\pm 66$&0.11\\
PMN J0210-5101$^*$& CMB6 & $3198$ &$1268\pm 86$&-0.28\\
PMN J0214-5054 & CMB6 & $61$ & $186\pm 63$&0.31\\
PMN J2235-4835$^*$& CMB7 &$1104$&$656\pm 63$&-0.16\\
\hline
\end{tabular}
\tablecomments{\small
These sources from the PMN $4.85\,$GHz catalog are detected at 
$>2.8\sigma$ significance
with ACBAR, corresponding to a false detection rate of 1. 
The fluxes at $4.85\,$GHz ($S_{4.85}$, from \citet{wright94}) and $150\,$GHz ($S_{150}$,
measured by ACBAR) are given. 
The spectral index $\alpha$ is defined as $S_\nu\propto \nu^{\alpha}$. 
The uncertainties associated with $S_{150}$ are dominated by
the CMB fluctuations. The central guiding quasars (one in each of the 4 fields) 
are marked with asterisks ($^*$).
These sources, as well as the undetected PMN sources, are projected out from the 
data using the methods described by K04 and do not contribute to the 
power spectrum measurements reported in this paper.
}
\label{tab:pmnsources}
\normalsize
\end{center}
\end{table*}

\section{Results And Discussions}\label{sec:results}

\subsection{Power Spectrum}

\begin{figure*}[ht!]
\resizebox{\hsize}{!}{
\plotone{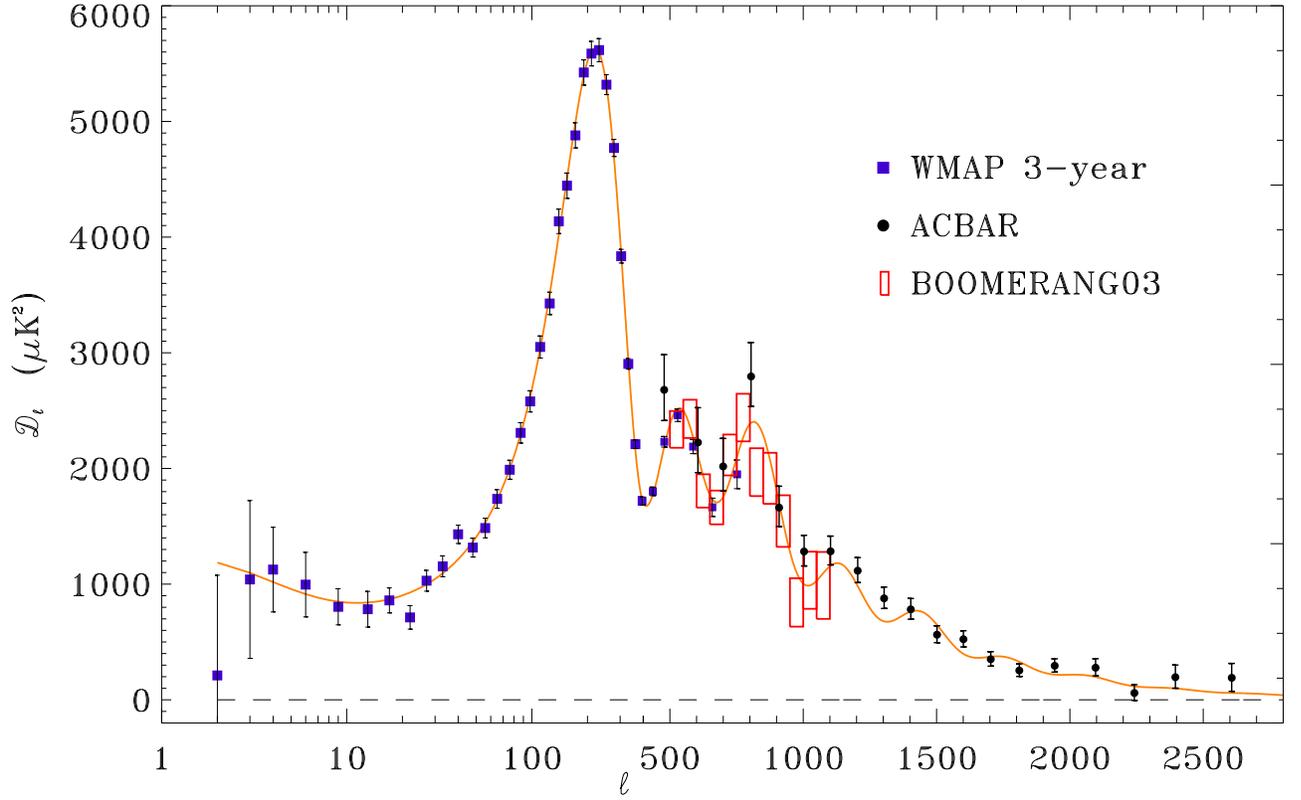}}
\caption{\protect\small
The ACBAR band powers plotted with those from WMAP3 \citep{hinshaw06} 
and the 2003 flight of BOOMERANG \citep{jones06}. The three 
experiments show excellent agreement in the overlapped region.}
\label{fig:acbar_ext}
\end{figure*}

Applying the analysis method described in the previous sections 
to the ACBAR 2001 \& 2002 $150\,$GHz data leads to the power spectrum shown in 
Figure~\ref{fig:acbar}. 
A comparison with other recent CMB results is shown 
in Figure~\ref{fig:acbar_ext}. The model curves in both figures are 
the ``WMAP3+ACBAR" best fit model.
We report the decorrelated band-powers, since
the description of their statistical properties requires fewer parameters. 
Plotting the decorrelated band-powers also simplifies the visual comparison 
between models and the measurements. 
The decorrelation transformations are defined according to \citet{tegmark97b}.
The same transformations are applied to the window functions, which convert 
a model $C_{\ell}$ to the theoretical band-powers \citep{knox99}.   
Following K04, we use the offset lognormal
functions \citep{bond2000} to fit the likelihood functions, and report
the fit parameters ${\bf q},{\bf \sigma},{\bf x}$ for each band.
The band-powers, uncertainties, and lognormal offsets are given in 
Table~\ref{tab:bands}; this information along with the
corresponding window functions are available for download from the ACBAR website 
\footnote{http://cosmology.berkeley.edu/group/swlh/acbar/index.html}.
 
The ACBAR data are consistent with the results of other CMB experiments, 
and fit the model predictions for a flat, $\Lambda$-dominated 
universe with a low baryon density.
A narrow peak is clearly seen in the power spectrum at 
$\ell \sim 820$, corresponding to the third harmonic of the acoustic 
oscillations in the early universe. This detection is in agreement with previous 
detections of this feature
by BOOMERANG \citep{jones06} and further confirms the coherent origin of 
the cosmic perturbations \citep{albrecht96}.  
Despite the nearly exponential damping at the high-$\ell$, 
the primary CMB fluctuations are detected with a signal-to-noise ratio of 
greater than $4$ up to $\ell=2000$. The photon diffusion mechanism
predicted at high-$\ell$ is verified to a high degree of accuracy.

\subsection{Anisotropies at $\ell>2000$} \label{sec:excess}

Various theoretical models predict that the secondary anisotropy 
induced by the thermal SZ effect in clusters of galaxies begins to dominate 
over primary anisotropy at $\ell \gtrsim 2000$ for standard cosmological parameters
\citep{cooray00,komatsu02}. 
The level of the signal is extremely sensitive to the normalization of the the matter 
power spectrum, usually parameterized by the present-day RMS mass fluctuation on 
8 $h^{-1}$ Mpc scales, $\sigma_8$.  The SZ effect has a clear spectral signature.
In the nonrelativistic limit, the thermodynamic temperature difference from SZ effect 
($\Delta T_{SZ}$) is given by
\begin{equation}
\frac{\Delta T_{SZ}}{ T_{CMB}} = y \left(x\frac{e^x + 1}{e^x - 1} - 4\right),\,\label{szs}
 \end{equation}
where $x = \frac{h\nu}{ k T_{CMB}}$ = $\nu/56.8\,$GHz. The quantity $y$ is known as the Compton
parameter and is proportional to the integrated electron density along the line of sight 
through the cluster ({\em e.g.}, \citet{peacock99}).
The CBI Deep field observations at $30\,$GHz detect power on scales corresponding to 
$\ell>2000$ in excess of the predicted primary CMB anisotropy.
This ``excess power" has been interpreted as the SZ effect produced by 
intervening galaxy clusters \citep{mason03,readhead04,bond02}. 
On the other hand, a variety of models including 
non-standard primordial effects have also been proposed as possible explanations 
\citep{voids,Bfields}. 
The unique photon emission spectrum of the thermal SZ effect distinguishes it from
these alternative explanations for the observed anisotropy.

\begin{figure*}[ht!]
\resizebox{\hsize}{!}{
\plottwo{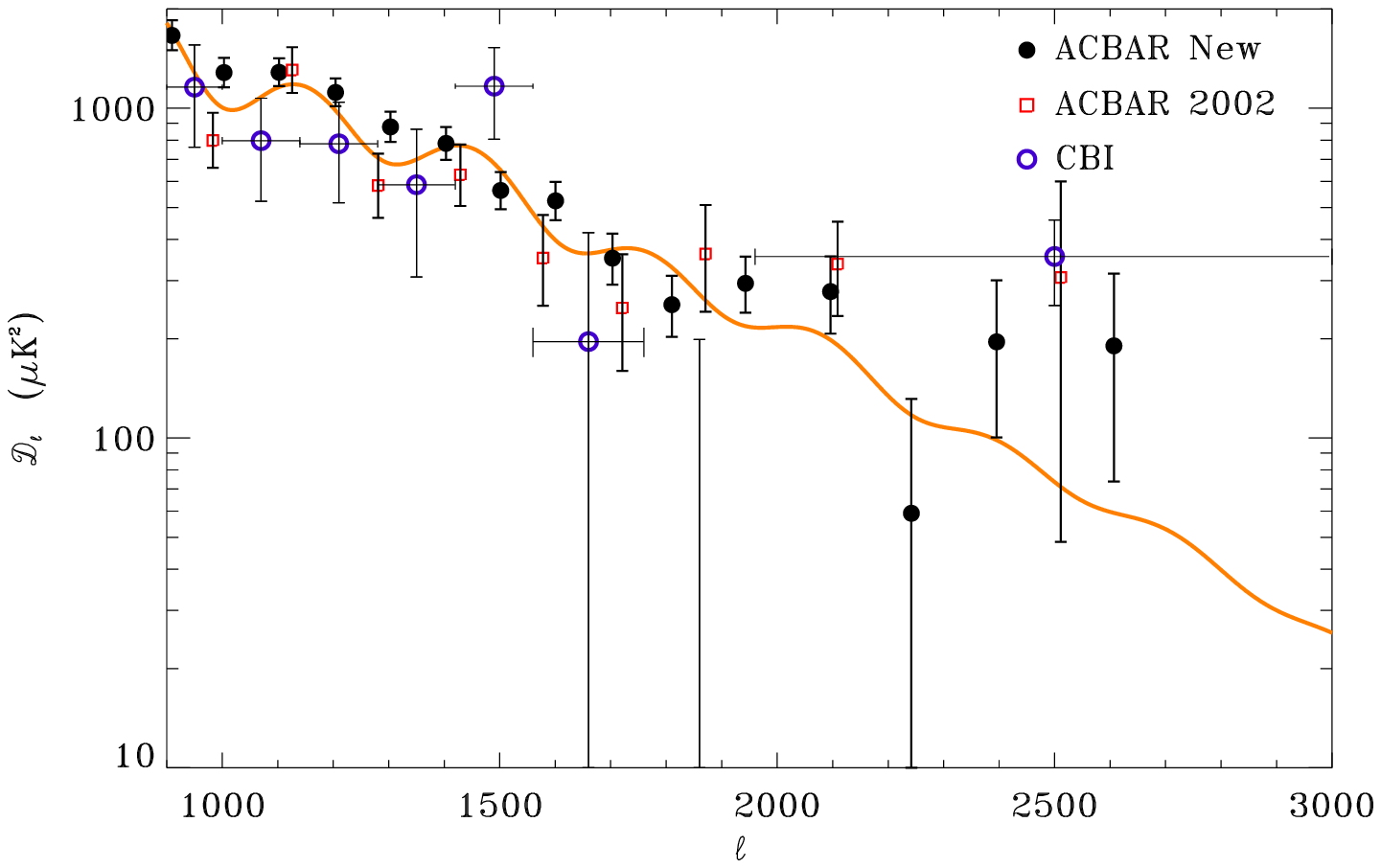}{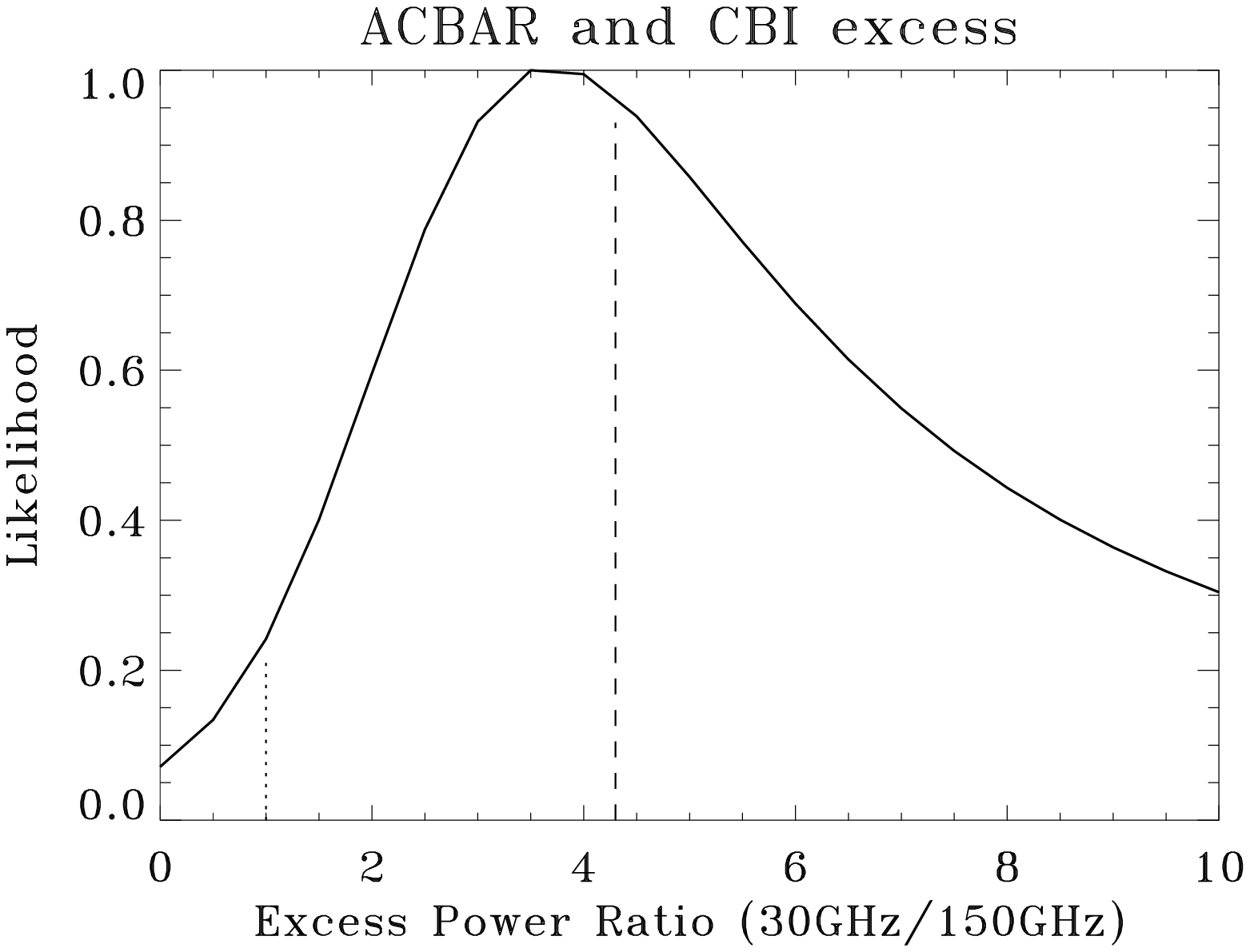}}
\caption{ACBAR results on the high-$\ell$ anisotropies. {\em Left}:
The ACBAR band-powers at $\ell >1000$, plotted on a logarithmic scale 
with the latest CBI data taken at a frequency of 
$30\,$GHz. All the ACBAR bins at $\ell >2000$ are lower than the CBI band-power measurement. {\em Right}:
The likelihood distribution for the ratio of the ``excess" power, observed by CBI at 
$30\,$GHz and
ACBAR at $150\,$GHz. The excess for each experiment is defined as the difference of
the measured band-powers and the model band-powers at $\ell>2000$. 
The vertical dashed line represents the
expected ratio (4.3) for the excess being due to the SZ effect. 
If the excess power seen in CBI is caused 
by non-standard primordial processes, the ratio will be unity (blackbody),
indicated by the dotted line. 
We conclude that it is 4.5 times more likely that the excess seen by CBI and 
ACBAR is caused by the thermal SZ effect than a primordial source.
In addition, because of the weak detection of excess power in ACBAR ($1.2\sigma$), 
it is about 3 times more likely that the excess is due to the SZ effect than
radio source contamination of the lower frequency CBI data, assuming no contaminations from dusty proto-galaxies. 
}
\label{fig:excess}
\end{figure*}

The ACBAR band powers corresponding to the smallest angular scales lie slightly above 
the best fit WMAP3 $\Lambda$CDM model. 
The four highest $\ell$ bins jointly produce an excess of $51\pm 42 \mu$K$^2$ after the 
model primary power spectrum is subtracted. 
The combination of this result with measurements at lower frequencies can be used to
constrain the photon emission spectrum of the excess, shedding light on its origin. 
We perform a joint analysis of the CBI results and the new ACBAR 
data at $\ell>2000$, assuming the contributions from primary 
anisotropy are known.
In each experiment, the theoretical band-powers for primary anisotropy 
are calculated from the product of the $\Lambda$CDM power spectrum 
and band window functions, which are then subtracted 
from the observed band-powers. 
A two-dimensional likelihood function is calculated from these excess band-powers and their 
uncertainties, where the two parameters are the ratio of the $30\,$GHz and 
$150\,$GHz excess, $\zeta$, and the power at $30\,$GHz, $\sigma^2_{30}$ (in $\mu K_{CMB}^2$).    
We then marginalize over the $\sigma^2_{30}$ parameter and plot the likelihood function for 
the power ratio $\zeta$ in Figure~\ref{fig:excess}. 
Since ACBAR measures significantly less power at $150\,$GHz, the data disfavor
sources that result in a blackbody spectrum ({\it i.e.}, $\zeta=1$). 
Using the ACBAR and CBI frequency response and equation 
(\ref{szs}), we calculate the power ratio $\zeta=4.3$ for the thermal SZ effect.
From the likelihood plot, we conclude that it is 4.5 times more likely that the excess 
seen by CBI and ACBAR is the result of the thermal SZ effect ($\zeta=4.3$) than 
a primordial process ($\zeta=1$). Since the expected ratio of flux at $30\,$GHz to 
$150\,$GHz from radio sources is expected to be $<$0.1, such sources are disfavored 
as being responsible for the excess power seen by CBI at about the same significance 
($\sim 1.2 \sigma$) with which ACBAR detects excess power.
Additional data from ACBAR or other higher frequency instruments will be required to
make a definitive statement about the origin of the excess power seen by CBI and BIMA.  

\section{Cosmological Parameters}\label{sec:parameters}

In this section, we estimate cosmological parameters for a minimal
inflation-based, spatially-flat, tilted, gravitationally lensed, $\Lambda$CDM model
characterized by six parameters, and then investigate models including
extra parameters to test extensions of the theory.  For
our base model, the six parameters are: the physical density of
baryonic and dark matter, $\Omega_bh^2$ and $\Omega_ch^2$; a constant
spectral index $n_s$ and amplitude $\ln A_s$ of the primordial power
spectrum, the optical depth to last scattering, $\tau$; and the ratio
of the sound horizon at last scattering to the angular diameter
distance, $\theta$.  The primordial comoving scalar curvature power
spectrum is expressed as ${\cal P}_s(k) = A_s (k/k_n)^{(n_s-1)}$,
where the normalization (pivot-point) wavenumber is chosen to be $k_n
= 0.05\, {\rm Mpc}^{-1}$.  The parameter $\theta$ maps angles observed
at our location to comoving spatial scales at recombination; changing
$\theta$ shifts the entire acoustic peak/valley and damping pattern of
the CMB power spectra.  Additional parameters are derived from the
basic set. These include: the energy density of a cosmological
constant in units of the critical density, $\Omega_\Lambda$; the age
of universe; the energy density of non-relativistic matter,
$\Omega_m$; the {\it rms} (linear) matter fluctuation in $8h^{-1}$Mpc
spheres, $\sigma_8$; the redshift to reionization, $z_{re}$; and the
value of the present day Hubble constant, $H_0$, in units of
kms$^{-1}$Mpc$^{-1}$.  
Tilted primordial spectra indicate the presence of a  
tensor-induced anisotropy component, however, we do not include this potential
contribution due to its uncertain amplitude.
The influence of the tensor component would only be significant
at low-$\ell$, not in the regime which ACBAR probes. 
We also restrict this work to flat
$\Lambda$CDM models, motivated by the observed curvature being so close to
zero. However, we have run models with non-zero curvature $\Omega_K$, and find that they
reproduce the standard geometrical degeneracy associated with $\Omega_K$ and
$\Omega_\Lambda$.

We have also considered two extensions to the basic model which could potentially 
impact the interpretation of the ACBAR bandpowers. 
These extended models include flat $\Lambda$CDM models with a
running scalar spectral index characterized by the derivative
$dn_s/d\ln k (k_n)$, and flat $\Lambda$CDM models with  
a Sunyaev-Zel'dovich contribution to the angular power spectrum
with amplitude parametrized by $\alpha^{\rm SZ}$. 
We also investigate a model where both a running spectral index and
a SZ contribution are considered simultaneously. 

The parameter constraints are obtained using a Monte Carlo Markov
Chain (MCMC) sampling of the multi-dimensional likelihood as a
function of model parameters. Our software is based on the publicly
available {\textsc CosmoMC}\footnote{http://cosmologist.info/cosmomc}
package \citep{Lewis:2002ah}. CMB angular power spectra and matter
power spectra are computed using the {\textsc CAMB} code
\citep{lewis00}. We approximate the full non-Gaussian bandpower likelihoods
with an offset lognormal distribution \citep{bond2000} found by explicit 
fits (see K04 for a detailed discussion of the calculation).  Our standard
{\textsc CosmoMC} results include the effects of weak gravitational
lensing on the CMB \citep{seljak96,lewis00}. Lensing effects in the
temperature spectrum are expected to become significant at
scales $\ell > 1000$, hence it is important to include this effect
when interpreting the ACBAR results.  The major effect of
lensing is a scale-dependent smoothing of the angular power spectrum
which diminishes the peaks and valleys of the spectrum. 
Inclusion of lensing in the model improves the fit to the data for all
experiment combinations.
However, we find that the parameter mean values and uncertainties are 
largely unaffected by
the inclusion of lensing with some exceptions, in particular
the introduction of lensing tends to increase the value of $\sigma_8$.

The typical computation consists of $8$ separate chains, each having
different initial random parameter choices. The chains are run until
the largest eigenvalue of the Gelman-Rubin test is smaller than 0.1
after accounting for burn-in. Uniform priors with very broad
distributions are assumed for the basic parameters. The standard run
also includes a weak prior on the Hubble constant ($45 < H_0 < 90$
km\, s$^{-1}$\, Mpc$^{-1}$) and on the age of the universe ($>10$
Gyrs). We also investigate the influence of adding Large Scale Structure
(LSS) data from the 2 degree Field Galaxy Redshift Survey (2dFGRS)
\citep{cole05} and the Sloan Digital Sky Survey (SDSS)
\citep{tegmark04}. When including the LSS data, we use only the band powers
for length scales larger than $k \sim 0.1 h$Mpc$^{-1}$ to avoid non-linear
clustering and scale-dependent galaxy biasing effects. We marginalize
over a parameter $b^2_g$ which describes the (linear) biasing of the
galaxy-galaxy power spectrum for $L_\star$ galaxies relative to the
underlying mass density power spectrum. We adopt a Gaussian prior on
$b^2_g$ centered around $b_g=1.0$ with a conservative width
equivalent to $\delta b_g = 0.3$; all parameters except $\sigma_8$ and
$\tau$ are insensitive to this width.

\subsection {Base Parameters Results}\label{sec:basic}

The results for the basic flat tilted $\Lambda$CDM parameters are
shown in Table~\ref{tab:basic}. The confidence limits are obtained by
marginalizing the multi-dimensional likelihoods down to one
dimension. The median value is obtained by finding the 50\% integral
of the resulting likelihood function while the lower and upper error limits are
obtained by finding the 16\% and 84\% integrals respectively. The
CMBall data combination includes: the ACBAR results presented here; the
WMAP 3 year TT, TE, and EE spectra, with the EE not included at higher
$\ell$ as in \citet{hinshaw06}; the CBI extended mosaic results
\citep{readhead04} and polarization results \citep{Readhead04b,Sievers05}, 
combined in the manner described in 
\citet{Sievers05};\footnote{We exclude the band powers below $\ell=600$ from
the CBI extended mosaic results to reduce the correlation with the TT
band powers of the CBI polarization dataset which influence the
sample-dominated end of the spectrum.} the DASI two year results
\citep{halverson02}; the DASI EE and TE bandpowers \citep{Leitch04}; the VSA final results
\citep{dickinson04}; the MAXIMA 1998 flight results
\citep{hanany00}; and the TT, TE, and EE results from the BOOMERANG
2003 flight \citep{jones06, piacentini06, montroy06}.  Only $\ell > 350$ bandpowers are
included for BOOMERANG because of overlap with WMAP3 (although
inclusion of the lower $\ell$ results leaves the parameter results
essentially unchanged).  While ACBAR and BOOMERANG are both
calibrated through WMAP, this is a small contribution to the total
uncertainty in the ACBAR calibration and we treat the calibration
uncertainties as independent in our parameter analysis. Although the
DASI, CBI and BOOMERANG 2003 EE and TE results for high $\ell$
polarization are included, they have little impact on the values of
the parameters we obtain.  

In all our runs we have used the updated WMAP3 likelihood code
(http://lambda.gsfc.nasa.gov/) which includes an updated point-source
correction {\sl cf.} \citet{huffenberger06} and foreground
marginalization on large angular scales. These updates result in 
a small increase in the $\Omega_m$ and $\sigma_8$ values compared to those
reported in \citet{spergel06}.

\begin{figure*}[ht!]
\centering
\includegraphics[width=5.5in]{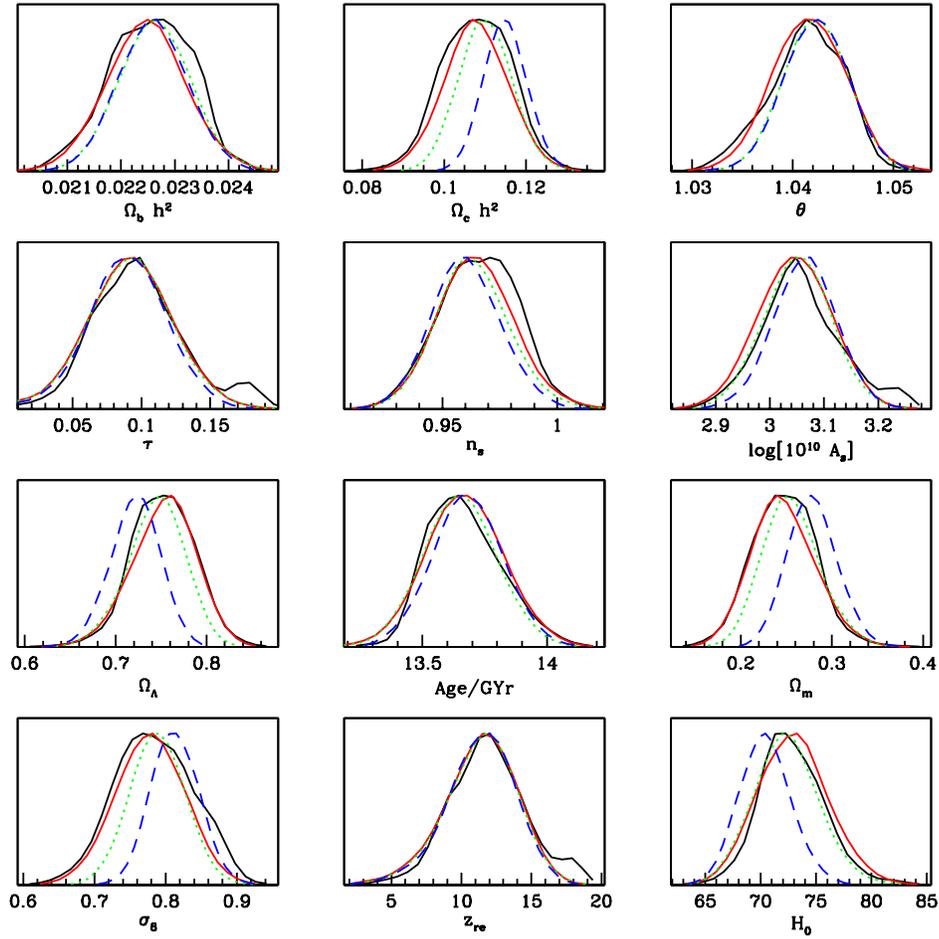}
\caption{Basic parameter marginalized 1-dimensional likelihood
distributions for the following data combinations; WMAP3 only (black,
solid), ACBAR + WMAP3 (red, dashed), CMBall (green,
long-dashed), and CMBall + LSS (blue, dash-dot). All runs
include lensing. 
}
\label{fig:basic}
\end{figure*}

The results for the basic model parameter set with various
combinations of data are summarized in Fig.~\ref{fig:basic}.  The most
striking feature of the results is that the solutions determined from
WMAP3 alone are quite compatible with the extension by ACBAR (and that
of the other data) to higher $\ell$.  This consistency means that the
additional CMB data (including ACBAR) have little impact on the
cosmological parameters determined by WMAP3. 
We have tested the effect of a significantly smaller ACBAR calibration error, 
such as we anticipate for the final ACBAR release. We find
a much larger impact on the parameter values and errors;
the values are similar to those found for CMBall+LSS.

With the original \citet{spergel06} WMAP3 likelihood code,
there was a shift in $\sigma_8$ and $\Omega_m$ to higher values when
additional data was included. However, with the updated WMAP3 likelihood
code, the addition of the ACBAR and CMBall bandpowers leads to
essentially no shift in $\sigma_8$ and $\Omega_m$; however, including
the LSS data does still result in a slight increase in these parameters. 
The new likelihood code
corrects the lower power in the third acoustic peak which was leading 
to low values for  $\sigma_8$ and $\Omega_m h^2$. 

The comoving damping scale, determined as a derived
cosmological parameter using only the ACBAR and WMAP3 data is $R_D=
10.5  \pm 0.2 \, {\rm Mpc}^{-1}$. The corresponding angular scale is
$\ell_D= 1355 ^{+5}_{-5}$. 
These
values for $R_D$ and $\ell_D$ are in excellent agreement with values
obtained using earlier datasets \citep{BCP03}. We also find the
comoving sound crossing distance is $R_s= 147.8^{+2.3}_{-2.3} \, {\rm
Mpc}^{-1}$, with a corresponding angular scale $\ell_s= 100/\theta =
95.9^{+1.0}_{-0.2}$, in agreement with the value for $\theta$ in
Table~\ref{tab:basic}.

Inclusion of lensing in our standard parameter runs increases the
best-fit model likelihoods in all cases. The difference between the log
likelihoods of the lensed and non--lensed models for the WMAP3 run is
$\Delta \ln L=0.86$. The log likelihood difference increases
to 1.7 with ACBAR included, 2.46 with CMBall, and 3.69 for the CMBall+LSS data
combination. 
The mean values of the parameters do not
shift significantly with the inclusion of lensing; for example, $\sigma_8$ increases from 0.778
to 0.788 for the CMBall data set and from 0.804 to 0.813 for CMBall+LSS. 
The best-fit ${\cal D}_\ell$'s for the lens and no-lens cases look quite similar,
but the subtle smoothing of the peaks and troughs by lensing results in a better fit to the
the data for each combination of experiments.

\subsection{Running Spectral Index}\label{sec:nrun}

The first release of WMAP data showed evidence for running of the CMB power spectrum
spectral index, particularly when combined with measurements of LSS \citep{spergel03}.
Extending the basic model to allow for running of the spectral index
around the pivot point $k_\star=0.05$Mpc$^{-1}$ yields $dn_s/d\ln k
(k_\star)=-0.053^{+0.031}_{-0.029}$ for WMAP3 only. The tendency for
negative running indices is due mostly to the low $\ell$ end, where
the multipoles are lower than the standard $\Lambda$CDM model.  The
contribution from the high $\ell$ end is less significant.  Since the WMAP3
results extend to reasonably high $\ell$, the addition of the ACBAR
results shifts the constraints only marginally $dn_s/d\ln k (k_\star)=
-0.045^{+0.026}_{-0.026}$. The effect of adding the ACBAR data can be
seen most clearly in Fig.~\ref{fig:nrun} which shows the correlation
between $n_s$ and $dn_s/d\ln k$. The central value is similar, but the
errors are further reduced with the CMBall + LSS combination, 
$dn_s/d\ln k (k_\star)= -0.047^{+0.021}_{-0.021}$. 
Similar to the results from WMAP1 and earlier versions of the CMBall data set
\citep{BCP03,mactavish05}, a negative running is still favored at about the 
2-$\sigma$ level by each of the data combinations considered.
The models including running favor
significantly lower values of the scalar spectral index,
$n_s=0.903^{+0.029}_{-0.028}$. However this result depends on the choice
of pivot point $k_\star$: a smaller value would yield a higher result
while a higher one would give an even lower result. 

\subsection{Sunyaev-Zel'dovich template extension}\label{sec:sz}

As described in Section ~\ref{sec:excess}, fluctuations from the 
thermal Sunyaev-Zel'dovich (SZ) effect are
expected to dominate over the damped primordial contributions to the
CMB anisotropy at multipoles beyond $\ell \sim 2500$. 
The magnitude of the SZ signal depends strongly on the overall matter
fluctuation amplitude, $\sigma_8$.
We have
modified our parameter fitting pipeline to allow for extra frequency
dependent contributions to the CMB power spectrum and have implemented
it in a simple analysis using a fixed template ${\hat {\cal
C}}_\ell^{\rm SZ}$ for the shape of the thermal SZ power spectrum. The
template was obtained from large hydrodynamical simulations of a
scale-invariant ($n_s=1$) $\Lambda$CDM model with $\sigma_8=0.9$ and
$\Omega_bh = 0.029$. (See \cite{bond05} for a detailed description of
the simulations.) Recently the WMAP team have used a different SZ template
based on analytic estimations of the power spectrum
\citep{spergel06}. It is characterized by a slower rise in $\ell$ than
the simulation-based one, which cut nearby clusters out of the power
spectrum. There has been no fine-tuning of the spectra to agree with
all of the X-ray and other cluster data. This may have an effect on
shape, especially at high $\ell$.

We add an SZ contribution ${\cal C}_\ell^{\rm SZ} = (\alpha^{\rm
SZ})^2{\hat f_\nu {\cal C}}_\ell^{\rm SZ}$ to the base six parameter
model spectrum.  Here $f_\nu$ is the frequency-dependent SZ pre-factor
and $\alpha^{\rm SZ} = \sigma_8^{7/2} (\Omega_bh/0.029)$ is a scaling
factor determined from hydrodynamical simulations. We consider two
cases: (1) the scaling parameter $\alpha^{\rm SZ}$ is slaved to
$\sigma_8^{7/2} (\Omega_bh/0.029)$; (2) $\alpha^{\rm SZ}$ is allowed
to float freely.
Including this SZ template with all parameters varying is
complementary to the analysis of \S~\ref{sec:excess} which directly compared
the residual CBI and ACBAR bandpowers in the excess region. In that
more restrictive analysis, the primary power spectrum is fixed and
$f_\nu$ is allowed to vary as well as a broad-band excess power. 
We found that the spectrum of the excess was compatible with the SZ
effect, motivating the SZ-restricted study considered here.

Regardless of the data combination, we find that including an SZ component 
in the model has little effect on the determination of the basic cosmological 
parameters. 
This can be seen by comparing Table~\ref{tab:SZ} with Table~\ref{tab:basic}. 
We also find that it has little effect, whether $\alpha^{\rm SZ}$ is related
to cosmic parameters through $\alpha^{\rm SZ} = \sigma_8^{7/2}
(\Omega_bh/0.029)$ or is allowed to float freely.  Note that the SZ
results break the $A_s e^{-2\tau}$ near-degeneracy (as does weak
lensing, though not as strongly.)

We begin with the combination of the ACBAR and WMAP3 data. 
When $\alpha^{\rm SZ}$ is allowed to float freely, we obtain $\alpha^{\rm SZ}=
0.57^{+0.20}_{-0.56}$. We can use the above definition to map the floating SZ amplitude 
parameter $\alpha^{\rm SZ}$ to a corresponding $\sigma_{8}^{({\rm SZ})}=0.84^{+0.15}_{-0.23}$.
The low significance of excess power in the ACBAR data results in weak 
constraints on $\sigma_8^{\rm SZ}$, particularly for the lower limit.  
We have not tabulated the results for the slaved $\alpha^{\rm SZ}$ case since it
results in extremely small changes in $\sigma_8$. 
For example, fits to the ACBAR+WMAP3 band powers give $\sigma_8 =
0.77^{+0.05}_{-0.05}$ when we include the SZ contribution in the model,
and $\sigma_8 = 0.78^{+0.05}_{-0.05}$ when we ignore it. 

When the high $\ell$ bandpowers of CBI and BIMA are included in the
analysis, there is a significant detection of excess power. Both the CBI
and BIMA bandpowers are from $30\,$GHz interferometric observations and have
higher $f_\nu$ values than ACBAR. For the slaved case, the errors
tighten slightly while the central value remains stationary with
$\sigma_{8}= 0.79^{+0.04}_{-0.04}$. For the floating case we find
$\alpha^{\rm SZ}= 0.79^{+0.06}_{-0.09}$ which maps to
$\sigma_{8}^{({\rm SZ})}=0.92^{+0.05}_{-0.06}$.

The CMBall + BIMA combination results in uncertainties for
$\sigma_{8}^{({\rm SZ})}$ which are comparable to those of $\sigma_8$.
A visual summary of the results is shown in Fig.~\ref{fig:SZ} where we
plot both $\sigma_{8}$ and $\sigma_{8}^{({\rm SZ})}$ against the
spectral index for a number of data combinations.  It is interesting
to note that the tension between $\sigma_{8}$ and $\sigma_{8}^{({\rm
SZ})}$ is relaxed by the inclusion of the LSS data which increases
the value of $\sigma_8$.
We caution, however, that the fit depends on the
SZ template shape and its extension into the higher $\ell$ regime
probed by BIMA.  This analysis assumes that no additional foreground
sources, such as dusty proto-galaxies, contribute significantly
to the observed anisotropy.
\footnote{We note
that the non-Gaussian nature of the SZ signal is included in the BIMA 
results but was not taken into account in the CBI analysis. 
The effect of the sample variance tends to open up the allowed range towards 
lower $\sigma_8$ values \citep{goldstein03,readhead04}.}
It is worth mentioning that the baseline lensed model for the CMB continues to be a 
better fit to the data regardless of the inclusion of the SZ extension. 
The decrease in log likelihood for the fit to the model including SZ, when lensing 
is taken into account, is $\Delta \ln L = 2.06$ for 
CMBall+BIMA and $\Delta \ln L = 2.38$ for CMBall+BIMA+LSS.
We also find that neglecting lensing in the model tends to increase the 
existing tension between the derived $\sigma_8$ and $\sigma_8^{SZ}$ values.

We find that including the SZ template and running spectral index simultaneously
selects a more negative running index due to the greater downturn in
high $\ell$ power. For example, with the CMBall + BIMA +
LSS combination we get $dn_s/d\ln k = -0.60^{+0.022}_{-0.021}$
compared with the no-SZ result $dn_s/d\ln k = -0.47^{+0.021}_{-0.021}$. 
In either case, the excess power leads to $\sigma_{8}^{({\rm SZ})} =
0.92^{+0.04}_{-0.06}$, virtually unchanged from the results for the 
model without running.

\section{Conclusions}\label{sec:conclusion}

We have measured the CMB angular power spectrum up to multipole values
of $\ell \sim 3000$ using the complete data set from the 2001 and 2002
ACBAR $150\,$GHz observations.  The data are analyzed with a refined
method in which the band-powers are calculated from un-differenced
maps.  We calibrate the data by comparing the flux of the galactic HII
region RCW38 in the overlapped region of ACBAR and BOOMERANG; the flux
scale of BOOMERANG is calibrated through comparison with WMAP.  The
new calibration is found to be consistent with the previous
planet-based calibration, but with uncertainty reduced from $10\%$ to
$6.0\%$ in temperature.

We have carried out various jackknife tests to show that the ACBAR
band-powers are not compromised by systematic errors.  In deriving the
power spectrum, we have projected out known foreground modes including
the FSD dust template and the PMN radio sources, however, these
projections have no significant effect on the final band-powers.  The
contribution to temperature fluctuations at $150\,$GHz from high
redshift dusty protogalaxies remains uncertain.  However,
extrapolating from recent observations near $270\,$GHz, we determine
that these sources are unlikely to contribute significantly to the
ACBAR band-powers.

The band-powers presented in Table \ref{tab:bands} are the most
sensitive measurements of the CMB temperature anisotropy to date in
the range of $1000\lesssim \ell \lesssim 3000$.  The power spectrum
continues to support a spatially flat $\Lambda$CDM cosmology, with a
low baryonic density.  Since the WMAP3 data now extends into the third
peak, the addition of the damping tail data from ACBAR results in only
minor changes in the values and uncertainties of the standard cosmological 
parameters.  
For all combinations of data we have considered beyond WMAP3 alone, 
the baseline lensed CMB model results in significantly better fits than models
neglecting lensing.

The ACBAR data has also been used to place interesting constraints on
secondary anisotropies.  The band-powers for $\ell >2000$ are
significantly smaller than those reported by the CBI and BIMA
experiments and provide only a suggestion ($1.2\sigma$) of excess power above
what is expected from primary CMB anisotropy (\S~\ref{sec:excess}).
The excess power has a frequency spectrum consistent with the thermal
SZ effect and inconsistent with thermal sources such as primary
anisotropy.  Because of the weak detection of excess power by ACBAR,
radio sources are slightly disfavored as the source of
the signal, but cannot be ruled out.  

Theoretical work suggests that
the thermal SZ effect should be the dominant source of secondary
anisotropy.
The expected amplitude of the thermal SZ effect is extremely sensitive
to $\sigma_8$.  Adding the SZ amplitude to our cosmological parameter
runs, we infer values for $\sigma_{8}^{({\rm SZ})}$ that are somewhat
higher than the $\sigma_8$ found from the standard parameter runs, but
consistent within the uncertainties.
This tension is further reduced when LSS data is included in the
parameter runs. 

The results presented here are derived from a subset of the total
ACBAR data set which is currently being analyzed.  The final ACBAR
power spectrum at $150\,$GHz will include 3.7 times more effective
integration time, 6.7 times more sky coverage, and a direct calibration 
from comparison with WMAP3. 
This will result in significantly reduced uncertainties across the 
entire power spectrum and improved constraints on standard cosmological 
parameters as well as secondary anisotropies.

\vspace{1cm}

The ACBAR program has been primarily supported by 
NSF office of polar programs grants OPP-8920223 and OPP-0091840.
This research used resources of the National Energy Research Scientific 
Computing Center, which is supported by 
the Office of Science of the U.S. Department of Energy under Contract 
No. DE-AC03-76SF00098. Chao-Lin Kuo acknowledges support from a NASA postdoctoral 
fellowship
and Marcus Runyan acknowledges support from a Fermi fellowship. 
Christian Reichardt acknowledges support from a National Science 
Foundation Graduate Research Fellowship.
We thank members of the BOOMERANG team, in particular Brendan Crill, Bill Jones, and 
Tom Montroy for providing access to the B03 data, the pipeline used to generate 
simulation maps, and assistance with its operation.  

\begin{table*}[ht!]
\begin{center}
\caption{\label{tab:bands}Joint Likelihood Band-powers}
\small
\begin{tabular}{ccccc}
\hline\hline
\rule[-2mm]{0mm}{6mm}
$\ell$ range&$l_{eff}$ &$q$ ($\mu{\rm K}^2$)& $\sigma$ ($\mu{\rm K}^2$) &
x ($\mu{\rm K}^2$)\\
\hline
     351-550&     428&    2680&      284&    -698\\
     551-650&     605&    2225&      280&    -218\\
     651-750&     700&    2018&      227&    -286\\
     751-850&     804&    2796&      276&    -509\\
     851-950&     910&    1662&      174&    -257\\
    951-1050&    1003&    1282&      132&    -111\\
    1051-1150&    1102&   1284&      124&    -146\\
    1151-1250&    1204&   1116&      108&     -40\\
    1251-1350&    1303&    877&       92&     -43\\
    1351-1450&    1403&    782&       89&      63\\
    1451-1550&    1502&    563&       73&     118\\
    1551-1650&    1601&    524&       70&     139\\
    1651-1750&    1703&    351&       62&     266\\
    1751-1875&    1810&    254&       54&     272\\
    1876-2025&    1943&    294&       57&     307\\
    2026-2175&    2096&    278&       74&     520\\
    2176-2325&    2242&     59&       68&     508\\
    2326-2500&    2395&    196&      100&     851\\
    2501-3000&    2607&    190&      120&    1625\\
\hline
\end{tabular}
\tablecomments{Band multipole range and weighted value $\ell_{eff}$, decorrelated band-powers $q_B$, 
uncertainty $\sigma_B$
, and log-normal offset $x_B$ from the
joint likelihood analysis of CMB2, CMB5, CMB6 and CMB7.
The PMN radio point source and IRAS dust foreground templates have been
projected out in this analysis.}
\normalsize
\end{center}
\end{table*}

\begin{table*}
\begin{center}
\caption{ Basic 6 Parameter Constraints\label{tab:basic}}
\small
\begin{tabular}{c||cccc}
\hline\hline
\rule[-2mm]{0mm}{6mm}
& WMAP3  & WMAP3 + ACBAR & CMBall & CMBall+LSS\\
$        \Omega_bh^2$ &$ 0.0226^{+0.0008}_{-0.0007} $&$ 0.0225^{+0.0007}_{-0.0007} $&$ 0.0226^{+0.0006}_{-0.0006} $&$ 0.0226^{+0.0006}_{-0.0006} $ \\
$        \Omega_ch^2$ &$ 0.108^{+0.008}_{-0.009} $   &$ 0.108^{+0.008}_{-0.007} $   &$ 0.110^{+0.006}_{-0.006} $   &$ 0.115^{+0.005}_{-0.005} $    \\
$          \theta$    &$ 1.042^{+0.004}_{-0.004} $   &$ 1.042^{+0.004}_{-0.003} $   &$ 1.042^{+0.003}_{-0.003} $   &$ 1.042^{+0.003}_{-0.003} $    \\
$            \tau$    &$ 0.097^{+0.012}_{-0.014} $   &$ 0.092^{+0.014}_{-0.014} $   &$ 0.092^{+0.013}_{-0.014} $   &$ 0.090^{+0.013}_{-0.013} $    \\
$             n_s$    &$ 0.966^{+0.017}_{-0.017} $   &$ 0.964^{+0.016}_{-0.015} $   &$ 0.963^{+0.016}_{-0.014} $   &$ 0.960^{+0.015}_{-0.014} $    \\
$    \log[10^{10}A_s]$&$ 3.05^{+0.08}_{-0.06} $      &$ 3.05^{+0.06}_{-0.07} $      &$ 3.05^{+0.06}_{-0.06} $      &$ 3.07^{+0.05}_{-0.06} $       \\
$  \Omega_\Lambda$    &$ 0.75^{+0.03}_{-0.03} $      &$ 0.76^{+0.03}_{-0.04} $      &$ 0.75^{+0.03}_{-0.03} $      &$ 0.72^{+0.03}_{-0.03} $       \\
     Age [$Gyrs$]     &$ 13.7^{+ 0.2}_{- 0.1} $      &$ 13.7^{+ 0.2}_{- 0.1} $      &$ 13.7^{+ 0.1}_{- 0.1} $      &$ 13.7^{+ 0.1}_{- 0.1} $       \\
$          \Omega_m$    &$ 0.25^{+0.03}_{-0.03} $      &$ 0.24^{+0.04}_{-0.03} $      &$ 0.25^{+0.03}_{-0.03} $      &$ 0.28^{+0.03}_{-0.03} $       \\
$         \sigma_8$    &$ 0.78^{+0.06}_{-0.05} $      &$ 0.78^{+0.05}_{-0.05} $      &$ 0.79^{+0.04}_{-0.04} $      &$ 0.81^{+0.03}_{-0.03} $       \\
$        z_re$    &$ 11.8^{+ 2.7}_{- 2.6} $      &$ 11.5^{+ 2.3}_{- 2.6} $      &$ 11.6^{+ 2.2}_{- 2.6} $      &$ 11.5^{+ 2.2}_{- 2.4} $       \\
$        H_0$    &$ 72.7^{+ 3.0}_{- 2.3} $      &$ 72.9^{+ 3.1}_{- 3.1} $      &$ 72.3^{+ 2.8}_{- 2.8} $      &$ 70.4^{+ 2.3}_{- 2.2} $       \\
\hline
\end{tabular}
\tablecomments{Results for the basic parameter set. The runs all
assumed flat cosmologies, uniform and broad priors on each of the
basic six parameters and a weak prior on the Hubble constant ($45 <
H_0 < 90$ km\, s$^{-1}$ Mpc$^{-1}$) and the age ($> 10$ Gyr). All runs
included the effect of weak gravitational lensing on the spectra which
is significant for the $\ell > 1000$ scales probed
by ACBAR. The results are nearly identical if we include an SZ
template, with the largest, but still relatively minor, impact on
$\sigma_8$ and $\tau$.}
\end{center}
\end{table*} 

\begin{table*}
\begin{center}
\caption{SZ Template Parameter Constraints \label{tab:SZ}}
\small
\begin{tabular}{c||cccc}
\hline\hline
\rule[-2mm]{0mm}{6mm}
& ACBAR+WMAP3  & CMBall & CMBall+BIMA  & CMBall+BIMA+LSS\\
$        \Omega_bh^2$        &$ 0.0224^{+0.0008}_{-0.0006} $&$ 0.0226^{+0.0007}_{-0.0006} $&$ 0.0226^{+0.0006}_{-0.0006} $&$ 0.0226^{+0.0006}_{-0.0006} $   \\
$        \Omega_ch^2$        &$ 0.106^{+0.008}_{-0.007} $   &$ 0.108^{+0.007}_{-0.006} $   &$ 0.108^{+0.007}_{-0.006} $   &$ 0.114^{+0.005}_{-0.005} $      \\
$          \theta$           &$ 1.041^{+0.004}_{-0.004} $   &$ 1.042^{+0.003}_{-0.003} $   &$ 1.042^{+0.003}_{-0.003} $   &$ 1.042^{+0.003}_{-0.003} $      \\
$            \tau$           &$ 0.096^{+0.014}_{-0.014} $   &$ 0.092^{+0.015}_{-0.014} $   &$ 0.091^{+0.014}_{-0.014} $   &$ 0.090^{+0.012}_{-0.012} $      \\
$         \alpha^{\rm SZ}$   &$ 0.57^{+0.20}_{-0.56} $      &$ 0.82^{+0.14}_{-0.81} $      &$ 0.79^{+0.06}_{-0.09} $      &$ 0.77^{+0.07}_{-0.08} $         \\ 
$             n_s$           &$ 0.962^{+0.018}_{-0.016} $   &$ 0.960^{+0.016}_{-0.015} $   &$ 0.961^{+0.017}_{-0.014} $   &$ 0.958^{+0.015}_{-0.015} $      \\
$    \log[10^{10}A_s]$       &$ 3.05^{+0.06}_{-0.07} $      &$ 3.04^{+0.06}_{-0.06} $      &$ 3.04^{+0.07}_{-0.06} $      &$ 3.06^{+0.06}_{-0.05} $         \\
$  \Omega_\Lambda$           &$ 0.76^{+0.03}_{-0.04} $      &$ 0.75^{+0.03}_{-0.03} $      &$ 0.76^{+0.03}_{-0.03} $      &$ 0.73^{+0.02}_{-0.03} $         \\
     Age [$Gyrs$]            &$ 13.7^{+ 0.2}_{- 0.2} $      &$ 13.7^{+ 0.1}_{- 0.2} $      &$ 13.6^{+ 0.1}_{- 0.1} $      &$ 13.7^{+ 0.1}_{- 0.1} $         \\
$          \Omega_m$           &$ 0.24^{+0.04}_{-0.03} $      &$ 0.25^{+0.03}_{-0.03} $      &$ 0.24^{+0.03}_{-0.03} $      &$ 0.27^{+0.03}_{-0.02} $         \\
$             \sigma_8$           &$ 0.77^{+0.05}_{-0.05} $      &$ 0.78^{+0.04}_{-0.04} $      &$ 0.78^{+0.04}_{-0.04} $      &$ 0.81^{+0.03}_{-0.03} $         \\
$        z_{re}$           &$ 11.9^{+ 2.0}_{- 2.5} $      &$ 11.6^{+ 2.3}_{- 2.7} $      &$ 11.4^{+ 2.6}_{- 2.4} $      &$ 11.5^{+ 2.0}_{- 2.4} $         \\
$        \sigma_8^{{\rm SZ}}$           &$ 0.84^{+0.15}_{-0.23} $      &$ 0.92^{+0.08}_{-0.15} $      &$ 0.92^{+0.05}_{-0.06} $      &$ 0.90^{+0.05}_{-0.07} $         \\
$ H_0$       &$ 73.3^{+ 3.1}_{- 3.1} $      &$ 72.9^{+ 2.9}_{- 2.9} $      &$ 73.1^{+ 2.9}_{- 2.9} $      &$ 70.7^{+ 2.1}_{- 2.1} $         \\
\hline			            				  
\end{tabular}
\tablecomments{Marginalized parameter constraints for the SZ template runs with independent amplitude. The SZ template contribution is scaled by a frequency dependent factor for each experiment and an independent amplitude $\alpha^{\rm SZ}$. The parameter $ \sigma_8^{{\rm SZ}}$ is the normalization derived from $\alpha^{\rm SZ}$.}
\end{center}
\end{table*} 

\begin{figure*}[ht!]
\centering
\includegraphics[width=4.5in]{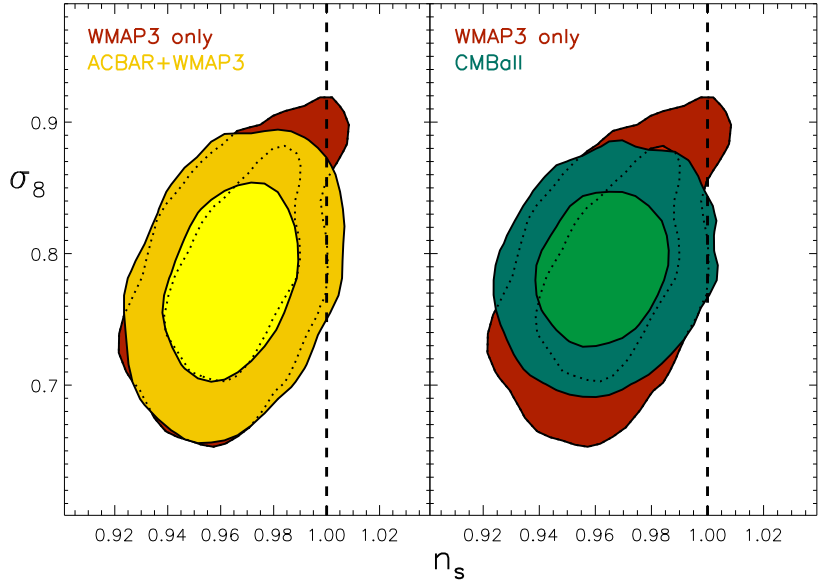}
\caption{Two-dimensional marginalized distribution for the $\sigma_8$
and the spectral index $n_s$. The contours are for the 68\%
and 95\% confidence levels. The underlying (red) region corresponds to 
the WMAP3 only basic parameter set in both left and right panels. In
the left panel the overplotted (yellow) region is for the combination
of ACBAR + WMAP3. In the right panel we show the combination ACBAR +
CMBall. The Harrison-Zel'dovich-Peebles, scale invariant solution
with $n_s=1$ is only ruled out at the 2$\sigma$ level. }
\label{fig:2d_nss8} 
\end{figure*}

\begin{figure*}[ht!]
\centering
\includegraphics[width=4.5in]{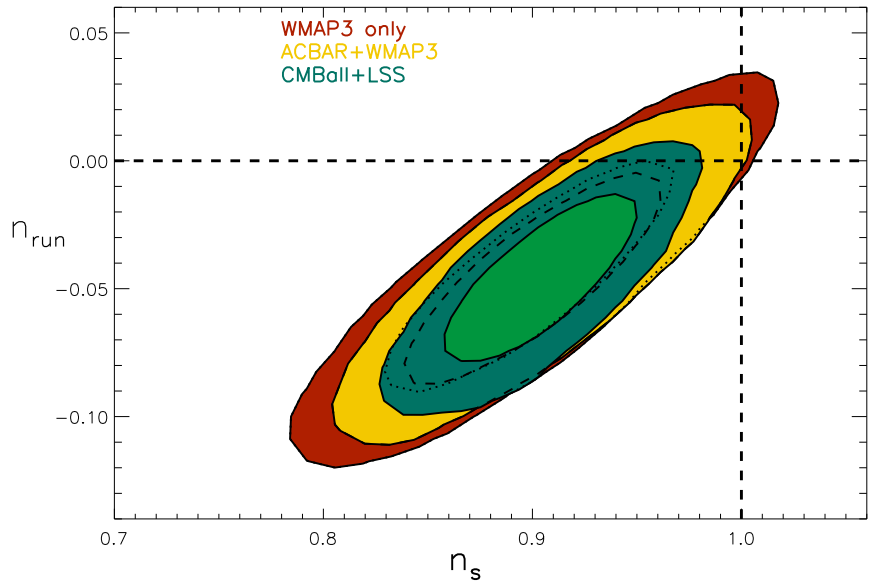}
\caption{Two-dimensional marginalized distribution for the
correlated pair $n_{run}=dn_s/d\ln k$ and $n_s$ for the three data
combinations. As in Fig. \ref{fig:2d_nss8}, the contours are for the 68\%
and 95\% confidence levels. The straight dashed lines show the
scale invariant case. This illustrates that negative running is
preferred at less than a 2-$\sigma$ level. If a tensor component was added,
the errors on running would become even larger.}
\label{fig:nrun} 
\end{figure*}

\begin{figure*}[ht!]
\centering
\includegraphics[width=4.5in]{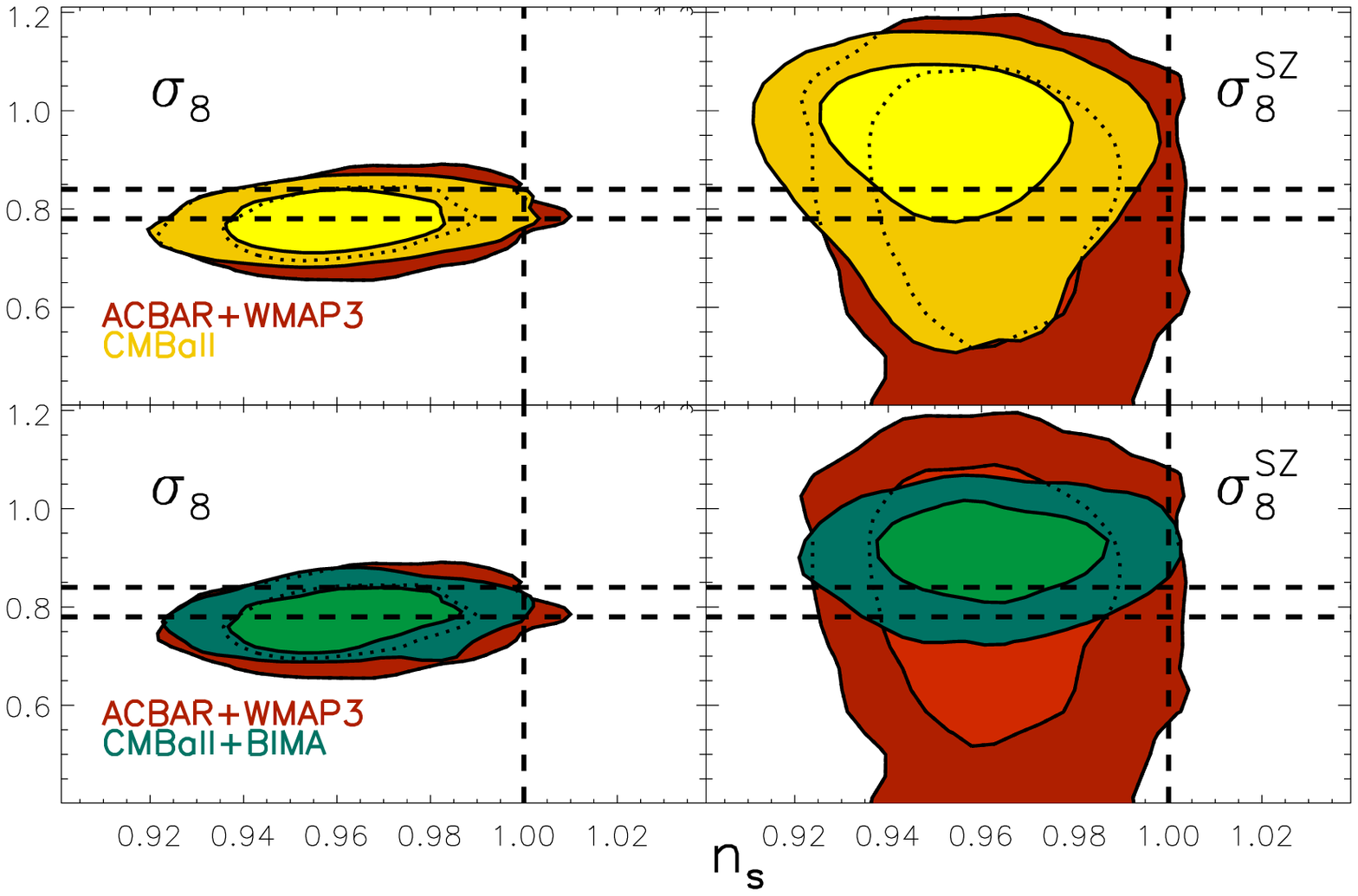}
\caption{Two-dimensional marginalized distribution for the
conventional $\sigma_8$ parameter (left panels) and 
SZ template fit derived $\sigma_8^{\rm SZ}$ (right panel), plotted {\sl vs.}
the spectral index $n_s$. The contours shown are for the 68\% and 95\%
confidence levels. The data combinations used are shown in the
insets. The horizontal lines show the 1$\sigma$ confidence region for
the $\sigma_8$ parameter obtained from the 1-d marginalized posterior
for the CMBall + BIMA + LSS combination with no SZ template fitting. 
The SZ derived values for the normalization of the matter fluctuations 
are higher than those found from the basic parameter run.}
\label{fig:SZ} 
\end{figure*}

\appendix

\section{A. CALIBRATION }\label{app:calib}

The calibration used for the first ACBAR data release was obtained
by observations of Venus and Mars, as detailed in \citet{runyan_the}. 
In this section, we describe a new calibration based on the comparison of
B03 and ACBAR observations of the galactic HII region RCW38.  
B03 was calibrated by an $a_{lm}$-based comparison of CMB structure with WMAP 
in the region of B03 sky coverage with a 1.8\% uncertainty~\citep{jones06}. 
The B03-ACBAR cross-calibration method is described below, with a detailed 
accounting of uncertainty in Table~\ref{tab:calerr}.

\paragraph{BOOMERANG-ACBAR Cross Calibration}

ACBAR made daily high signal-to-noise maps of the galactic source RCW38.  
B03 also mapped portions of the galactic plane including RCW38 (Fig. \ref{fig:rcw38}),
allowing a direct comparison of the high signal-to-noise maps made by the two 
experiments. The experiments have different scan patterns, beam widths, and spatial 
filters that can effect the measured flux. We resample the B03 map using pointing 
information for each ACBAR observation to generate an ACBAR-equivalent B03 
observation.  The ACBAR maps are smoothed to simulate the effect of Boomerang's 
larger beam. Large spatial modes in both experiments are corrupted; in ACBAR by 
chopper-synchronous offsets and in B03 by the high-pass filter.  We simultaneously 
fit a quadratic offset and Gaussian source model from each scan of the ACBAR 
and B03 maps which 
removes these modes without affecting the amplitude of a point source. 
After coadding the channel maps, we integrate the flux within a 18$^{\prime}$ 
radius of the source. The integrated flux is robust to small misestimates or 
changes in the beam size. The measured flux ratio and the associated uncertainties 
are listed under \textit{Ratio of B03 over ACBAR} in Table \ref{tab:calerr}.

We use Monte Carlo techniques to estimate the transfer function of this method.  
Using a model of RCW38 and its surroundings, we generate simulated 
timestreams for the observations with each experiment. Maps are created from the 
timestreams and are filtered as described above. The ratio of the transfer 
functions is found to be $ACBAR/B03 = 1.056 \pm 0.002$.  We have tested the 
dependence of the transfer function on the assumed signal template and include 
a 3\% uncertainty in our calibration due to this effect. This technique is readily 
adapted to include the effect of the beam uncertainty for each experiment and we 
find that the beam contributes 1.35\% to our estimated uncertainty.  
The effect of the transfer function and the associated uncertainty are listed 
under \textit{Transfer Function} in Table \ref{tab:calerr}. 

RCW38 has a much different spectrum than the CMB, and the effective CMB temperature 
difference it produces depends on the photon-frequency.  
The calibration described here is based on observations with ACBAR's $150\,$GHz 
channels and Boomerang's $145\,$GHz channels which have similar bandpasses 
(Fig. \ref{fig:rcw38}).  
We account for the small difference in bandpass by convolving the measured spectral 
response of each experiment with a model of RCW38's spectrum from \citep{masi05}.  
If two maps nominally calibrated in CMB temperature units are integrated about RCW38, 
the true calibration factor {\bf K } will depend on the measured flux ratio 
$I_{B03} / I_{ACBAR}$, the bandpass of each experiment $t_\nu$, the spectrum of 
RCW38 $S_\nu^{RCW38}$ and the known blackbody spectrum of the CMB 
$\frac{dB_\nu}{dT}|_{T_{CMB}}$:
\[ K = \frac{I_{B03}}{I_{ACBAR}} * R\]
\[ {\rm where} \  R = \frac{\int{t_\nu^{ACBAR} \lambda^2 S_\nu^{RCW38} d\nu}}{\int{t_\nu^{B03} \lambda^2 S_\nu^{RCW38} d\nu}} \frac{\int{t_\nu^{B03} \lambda^2 \frac{dB_\nu}{dT}|_{T_{CMB}} d\nu}}{\int{t_\nu^{ACBAR} \lambda^2 \frac{dB_\nu}{dT}|_{T_{CMB}} d\nu}} .\]

This factor R includes the full dependence of the calibration on RCW38's spectrum 
and the bandpasses of each experiment.  
The dominant source of uncertainty is RCW38's spectrum; to be conservative, 
we double the estimates listed in \citet{masi05}.
The model consists of two components: a power law term with $\alpha = 0.5 \pm 0.2$ 
and a dust term with $T_{dust} = 22.4 \pm 1.8\,$K.  
Only the relative amplitude of the two terms is important: 
$A_{power law @ 30\,GHz}/A_{dust peak} = 867 \pm 400$. 
We also include uncertainty in the laboratory measurement of each experiment's bandpass. 
The mean value and uncertainty in R is estimated using 100,000 realizations of the 
above parameters, and found to be $1.008 \pm 0.021$ (See \textit{Spectral Correction} 
in Table \ref{tab:calerr}.) Given that our integration radius is larger then 
RCW38's size, the flux contribution of diffuse emission near RCW38 can be significant. 
The spectrum of this extended structure may be different from that of RCW38, in 
which case the calibration ratio would depend on the integration radius. 
We estimate this uncertainty from the observed variability of the calibration 
ratio with integration distance.

The calibration value from the real map
is normalized by the spectral correction for RCW38 and the signal-only transfer 
functions estimated for each experiment. 
The result of this analysis is that the temperature scale for ACBAR's CMB fields in 
2002 should be multiplied by $1.128 \pm 0.066$ relative to the planet-based 
calibration given  in \citet{runyan03a}. 
Table~\ref{tab:calerr} tabulates the contributing factors and error budget. 
We now proceed to propagate this RCW38-based calibration to the CMB2 observation 
done in 2001.

\begin{table*}
\begin{center}
\caption{Error Budget for the RCW38 based ACBAR Calibration}
\small
\begin{tabular}{lcc}
\hline\hline
\rule[-2mm]{0mm}{6mm}
Source & Value & Uncertainty (\%)\\
\hline
Ratio of B03 over ACBAR  & 1.060 & - \\
\;\; Statistical error & & 0.53 \\
\;\; Residual chopper synchronous offsets &  & 0.1 \\
\;\; B03 Instrumental noise & & 0.3 \\
\;\; Variability during 2002 & & 2.0 \\
Transfer function: & 1.056 & -\\
\;\; Statistical error & & 0.17 \\
\;\; Uncertainty in the signal model  & &  3.0 \\
\;\; Dependence upon the radius of integration & &  1.5 \\
\;\;Beam uncertainty &  & 1.35\\
Spectral Correction & 1.008 & - \\
\;\; RCW38's spectrum and experimental bandpasses & & 2.1 \\
\;\; Spectrum of extended structure & & 3.0 \\

\hline
B03's Absolute Calibration through WMAP& & 1.8 \\

Overall & 1.128 & 5.84\% \\ 
\hline
\hline
\end{tabular}
\tablecomments{\small
The calibration of ACBAR through RCW38 has multiple factors and potential sources of error, 
tabulated here for reference.  
The dominant calibration uncertainties are due to uncertainties in the emission 
spectrum of RCW38 and the morphology and spectrum of the extended galactic structure. }
\label{tab:calerr}
\normalsize
\end{center}
\end{table*}

\begin{figure*}
\plotone{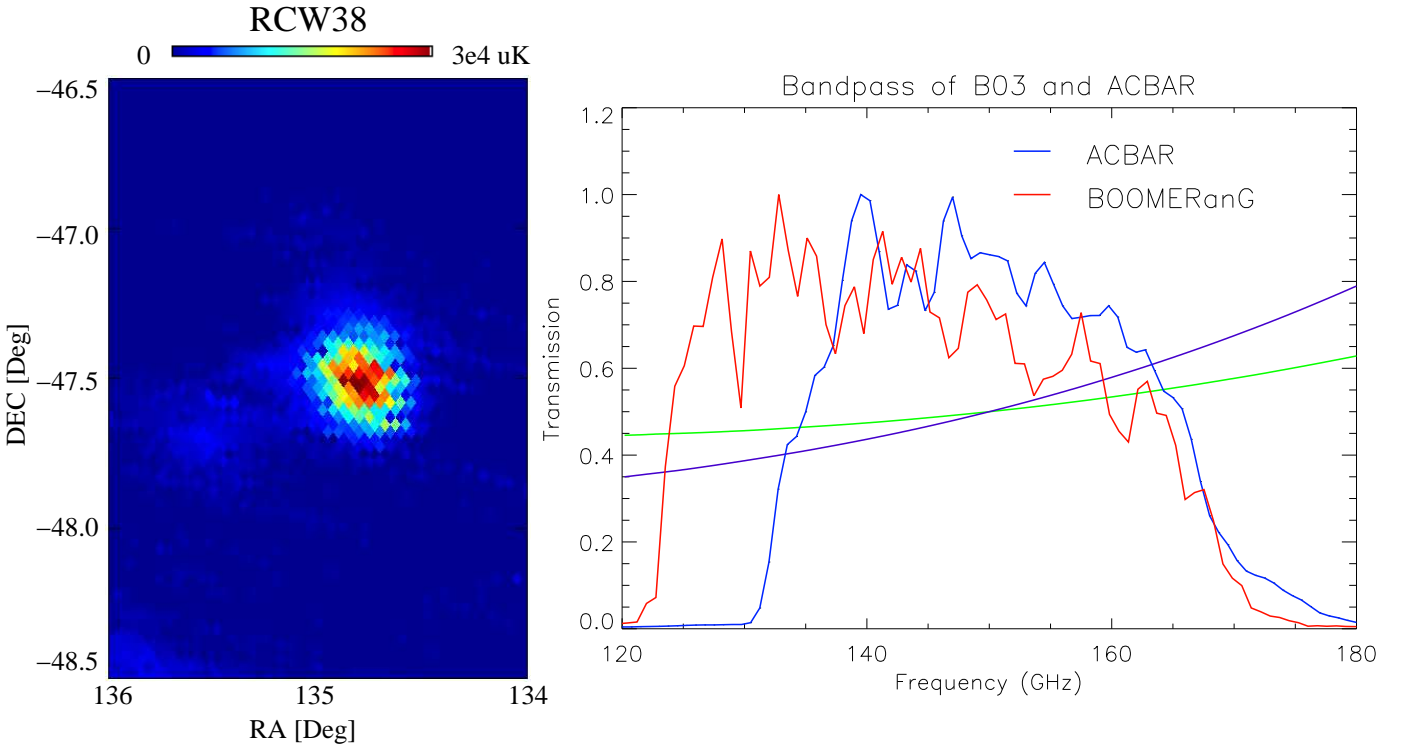}
\caption{On the {\bf left-hand side} is a map of RCW38 made by B03.  
The bandpasses for each experiment measured by Fourier transform spectroscopy are shown on the {\bf right}.  
The range of possible spectra (1$\sigma$) for RCW38 is also shown, with each spectrum being normalized to 
0.50 at $150\,$GHz. (green and violet lines).}
\end{figure*}\label{fig:rcw38}

\paragraph{ACBAR 2001-2002 Cross Calibration}

We carry the 2002 RCW38 calibration into 2001 by comparing the 
2001 observations of the CMB2 field to the overlapping 2002 CMB4 field.  
The fields are reduced to the overlapping region and a power spectrum is 
calculated for each field. The bands are widened ($\Delta \ell \sim 200$)
to avoid large noise correlations between the band-powers.
Care was taken to insure the filtering of the two maps only occurs in 
the overlapped region. 
However, differences in scan patterns and array 
configurations between the two seasons cause differences in filtering.
We assume that the band-powers from field $\alpha$ and field $\beta$ are
${\bf q}_f$ ($f=\alpha\,,\beta$). 
If the relative calibration factor between fields $\alpha$ and $\beta$ is $\eta$,
we can find the value $\eta_0$ that maximizes the likelihood function:
\begin{equation}
{\mathcal L}(\eta)\propto
\exp\left[-\frac{1}{2}\sum_i\frac{(q_{\alpha,i}-\eta q_{\beta,i})^2}
{\sigma_{D,i}^2}\right]\;.\label{calib_qb}
\end{equation}
The quantity $\sigma_{D,i}^2$ is the variance of 
$(q_{\alpha,i}-q_{\beta,i})$ for band $i$, which can be 
found by Monte Carlo analysis.
A separate Monte Carlo simulation was carried out to confirm that  
$\eta_0$ is an unbiased estimator, and to calculate its uncertainty.
We find the calibration factor to be CMB2/CMB4$= 1.238 \pm 0.067$ 
($\sqrt{\eta_0}$, in units of temperature). Approximating the uncertainties 
as Gaussian, it implies the CMB2 temperature scale should be multiplied by 
$0.911 \pm 0.074$ relative to the scale used for the analysis of K04.

The day-to-day relative calibration for the 2002 CMB fields is determined using 
the measured flux of RCW38.  The procedures used are outlined in more detail K04. 
During some parts of 2002, RCW38 observations are only available 
for one of the two rows of $150\,$GHz channels. 
We derive the relative calibration between two rows of bolometers during these 
periods using the CMB power spectrum comparison method described earlier. 
We find the corrections to the BOOMERANG-based calibration factors 
are $1.031\pm 0.025$, $0.935\pm 0.050$, and $0.998\pm 0.042$ for 
CMB5, CMB6, CMB7, respectively. We apply these corrections,
and determine the overall calibration uncertainty to be
$6\%$ (in temperature units) based on the uncertainties associated with
B03/ACBAR-2002 RCW38 cross calibration.

\bibliographystyle{apj}
\bibliography{merged}

\end{document}